\documentclass[onecolumn]{aa}
\usepackage{graphicx}
\begin{document}
%

  \title{Updated pulsation models for anomalous Cepheids}

  \author{M., Marconi\inst{1}, G., Fiorentino\inst{2,3}, F. Caputo\inst{3}}
   \offprints{}

   \institute{Osservatorio Astronomico di Capodimonte, Via Moiariello 16,
   I-80131 Napoli, Italy \and Universita' ``Tor Vergata'',
   via della Ricerca Scientifica 1, I-00133,  Roma, Italy
   \and
   Osservatorio Astronomico di Roma, Via di Frascati 33,
I-00040 Monteporzio Catone, Italy \\
              email: marcella@na.astro.it, giuliana@mporzio.astro.it, caputo@coma.mporzio.astro.it}






\date{Received .........; accepted ........}

\titlerunning{A theoretical investigation of ``anomalous'' Cepheids}

\authorrunning{Marconi et al.}

\abstract {A theoretical investigation of the pulsation behavior
of so-named ``anomalous'' Cepheids is presented. The study is
based on nonlinear convective pulsation models with $Z$=0.0001 and
0.0004, mass in the range 1.3-2.2$M_{\odot}$ and various
luminosity levels. Based on these computations, we derive period,
bolometric light curves and the edges of the instability strip,
showing that a variation of the metal abundance from $Z$=0.0001 to
0.0004 has quite small effects on these quantities. Then, using
bolometric corrections and color-temperature transformations, we
are able to provide the predicted relations connecting pulsational
properties (periods, amplitudes) with magnitudes and colors in the
various photometric bands. The theoretical pulsational scenario is
compared to observed anomalous Cepheids in dwarf spheroidal
galaxies and, in particular, the predicted mass-dependent
Period-Magnitude-Amplitude and Period-Magnitude-Color relations
are used to estimate individual mass values, as well as to
discriminate between fundamental (F) and first-overtone (FO)
pulsators.}

\maketitle

\section{Introduction}

Among the various types of radial pulsating stars, the so-named
``anomalous'' Cepheids (ACs) are, perhaps, the least explored
ones. They are rare in Galactic Globular Clusters, whereas they
have been observed in all the nearby dwarf spheroidal galaxies
(see e.g., Nemec et al. 1988, Nemec, Nemec \& Lutz 1994, Mateo
1998, Pritzl et al. 2002, Dall'Ora et al. 2003, Baldacci et al.
2003) in which they have been searched for. These variables are
intrinsically brighter than RR Lyrae stars and, when their
apparent magnitude is scaled to the magnitude level of RR Lyrae
stars observed in the same system, they arrange along a strip with
the less luminous ones ($\sim$ 0.5 mag brighter than RR Lyrae
stars) at periods $\sim$ 0.30 days and the more luminous ones
($\sim$ 2.0 mag brighter than RR Lyrae stars) at periods $\sim$
2.0 days. If compared with other luminous metal-poor variables
such as Population II Cepheids (see the recent review by
Wallerstein 2002), they appear significantly brighter, at fixed
period.

A general consensus exists in the literature (see, e.g., Bono et
al. 1997a [B97], Caputo 1998 and references therein) on the fact
that AC variables, like RR  Lyrae stars and Population II
Cepheids, are metal-poor He-burning stars in the post Zero Age
Horizontal Branch (ZAHB) evolutionary phase. As reviewed by Caputo
(1998), for low metal abundances ($Z\le$ 0.0004) and relatively
young ages ($\le$ 4 Gyr) the effective temperature of ZAHB models
reaches a minimum ($\log{T_e}\sim3.76$) for a mass of about
1.0-1.2 $M_{\odot}$, while if the mass increases above this value,
both the luminosity and the effective temperature start
increasing, forming the so called ``ZAHB turnover''.  For larger
metallicities, the more massive ZAHB structures have brighter
luminosities but effective temperatures rather close to the
minimum effective temperature. Within such an evolutionary
scenario, AC variables appear to belong to the post-turnover
portion of the ZAHB which crosses the instability strip at
luminosity higher than RR Lyrae pulsators. Their origin is still
debated and the most widely accepted interpretations are: (1) they
are young ($\le$ 5 Gyr) single stars due to recent star formation;
(2) they formed from mass transfer in binary systems as old as the
other stars in the same stellar system.

The general properties of observed ACs have been analyzed by
Nemec, Nemec \& Lutz (1994) and by Pritzl et al. (2002) who
provided empirical Period-Luminosity (PL) relations in the $B$ and
$V$ bands for both fundamental (F) and first overtone (FO)
pulsators. However, as shown by the quoted authors, the mode
assignment for these variables is not an easy task and
 the behavior in the PL, Period-Amplitude (PA) and
Period-Color (PC) planes does not always allow to discriminate
between F and FO pulsation modes.

Based on previous nonlinear convective pulsating models and
evolutionary tracks, B97 constrained the blue absolute magnitude
and the period of metal-poor central He-burning pulsators more
massive than RR Lyrae stars, finding a minimum mass for the
occurrence of this kind of variables around $\sim 1.3 M_{\odot}$
with $Z$=0.0001 and $\sim 1.8 M_{\odot}$ with $Z$=0.0004. The
theoretical limits of the distribution of these high mass
pulsators in the $\Delta{M_B}$ vs  $P$ plane, where $\Delta{M_B}$
is the calculated $M_B$ scaled to the ZAHB magnitude at the mean
effective temperature of the RR Lyrae strip, conformed well to the
observed data of most ACs and suggested a minimum mass of about
1.5 $M_{\odot}$. According to the position in the PL and PA
planes, B97 confirmed that observed ACs are a mix of F and FO
pulsators and derived two distinct non parallel $B$ band PL
relations  for the two pulsation modes. These relations were
significantly different from the empirical parallel ones derived
by NNL, whereas they better conform  with the analysis by Nemec et
al. (1988) and Pritzl et al. (2002).

However,  B97 pulsation models did not include the most recent
input physics and covered a restricted range of stellar
parameters. For this reason, we computed new models including
updated opacity evaluations  and spanning a wider range of stellar
parameters. In this paper we present
the results of these computations and some preliminary tests of
the predictive capability of the updated pulsational scenario.

The organization of the paper is the following: in Sect. 2 we
present the models, the results concerning the topology of the
instability strip and the light curves. The transformation from
theoretical to observational quantities is presented in Sect. 3,
together with a short discussion of the different types of mean
magnitudes and colors. In the same section, we give the relations
connecting pulsational parameters (periods, amplitudes) to
absolute magnitudes and colors. The comparison with observed ACs
in dwarf spheroidal galaxies is briefly discussed in Sect. 4 and
some concluding remarks close the paper.

\begin{table}
\centering \caption[]{Input parameters for the grid of AC
pulsation models with $Y$=0.24 and $Z$=0.0001 and
0.0004.\label{tab1}}

\begin{tabular}{lc}
\hline
\\
 $M/M_{\odot}$ & log $L/L_{\odot}$
\\\\
 \hline
\\
1.3 & 1.82, 1.92 \\
1.6 & 1.88, 1.98, 2.08 \\
1.9 & 1.98, 2.08, 2.18 \\
2.0 & 2.08, 2.18, 2.28 \\
2.2 & 2.08, 2.28 \\\\
\hline
\end{tabular}
\end{table}

\section{The pulsation models}

\subsection{Periods and instability strip}

The basic physics underlying radial pulsation, that is the
conversion of thermal energy into kinetic energy, is quite well
known (e.g., see the nice review by Madore \& Freedman 1991) and,
as early shown by Sandage (1958), the natural consequence of the
Ritter's period-mean density relation and of the Stefan's law
connecting luminosity, radius and effective temperature is that
the pulsation period is a function of the luminosity, mass and
effective temperature of the pulsator. Present models are computed
using a nonlinear, nonlocal and time-dependent convective
hydrodynamical code which has been already discussed in a series
of previous papers (see, e.g., Bono \& Stellingwerf 1994; Bono et
al. 1997a,b) and is not repeated here. It seems of importance only
to note that, with respect to linear envelope models, the
nonlinear approach provides many more predictions beside the
pulsation period, such as the luminosity variation along the
pulsation cycle, namely the bolometric light curve, and its
morphology. Moreover, the inclusion of convection, which tends to
quench pulsation, is fundamental to gain reliable information on
both the blue and red edges of the pulsation region. Even though
we are aware that further computations are needed to refine our
knowledge of the pulsation phenomenon, our hydrodynamical code and
physical assumptions have already shown a nice agreement with
several observed properties of classical Cepheid (see Bono,
Marconi \& Stellingwerf 1999; Bono, Castellani \& Marconi 2000b,
2002; Bono et al. 1999, 2002; Caputo et al. 2000, 2001; Fiorentino
et al 2002) and RR Lyrae models (see Bono, Castellani, Marconi
2000a, Bono et al. 1997b, 2003, Marconi et al. 2003), and we are
confident that they can be extended to other types of pulsating
stars in order to construct a homogeneous pulsational scenario. It
is also important to consider that the study of complex stellar
systems, where a variety of pulsating stars can be observed,
requires that the same input physics should be adopted for both 
pulsation and evolutionary computations. For these
reasons, as for RR Lyrae and classical Cepheid models, we adopt
the updated compilations of the opacity tables (Iglesias \& Rogers
1996) used in modern star model computations. Finally, we have
considered two chemical compositions representative of the
metal-poor abundances of most of the observed ACs ($Z$=0.0001 and
0.0004, with $Y$=0.24), a mixing-length parameter $l/H_p$=1.5, and
the mass and luminosity levels suggested by He-burning
evolutionary models (see Castellani \& Degl'Innocenti 1995), as
reported in Table 1.

For each mass and luminosity, the computations are performed by
decreasing the effective temperature $T_e$ by steps of 100 K, and
the period from the bluest to the reddest pulsating model is
derived\footnote{The entire set of results is available upon
request to the authors.}. Note that increasing (decreasing) by 100
K the effective temperature of the bluest (reddest) pulsators
yields non-pulsating structures. In analogy with other well
studied pulsators such as RR Lyrae stars (see van Albada \& Baker
1971, Bono et al. 1997b, Marconi et al. 2003) and classical
Cepheids (see Bono, Castellani \& Marconi 2000b), linear
regression through the results allows us to derive analytical
relations connecting the period of both fundamental (F) and
first-overtone (FO) modes to the intrinsic stellar parameters,
namely mass, luminosity, and effective temperature. We obtain

$$\log P_F=10.879(\pm 0.001)+0.818\log L-0.616\log M-3.309\log T_e\eqno(1a)$$

$$\log P_{FO}=10.198(\pm 0.001)+0.775\log L-0.530\log M-3.143\log T_e\eqno(1b)$$

\noindent where mass $M$ and luminosity $L$ are in solar units. As
a whole, the period of FO pulsators can be fundamentalised as
log$P_F$=log $P_{FO}$+0.13, in agreement with the procedure
currently adopted for RR Lyrae stars.

For fixed metal abundance, mass and luminosity, FO pulsators are
generally bluer than F pulsators, so that the blue edge of the
first-overtone pulsation region (FOBE) and the red edge of the
fundamental one (FRE) can be taken as representative of the
boundaries of the whole instability strip, at least in the studied
range of mass and luminosity. Giving the adopted steps of 100 K in
our computations, we adopt as FOBE and FRE the effective
temperature of the bluest FO model and of the reddest F model,
increased and decreased by 50 K, respectively, with an intrinsic
uncertainty of $\pm$ 50 K. As for the internal topology of the
pulsation region, the computations show that the blue edge for
fundamental pulsation (FBE) is bluer than the first-overtone red
edge (FORE), suggesting the occurrence of a middle region of
the instability strip where both the pulsation modes can be stable, similarly to what found for RR Lyrae stars (see van Albada \&
Baker 1973, Marconi et al. 2003 and references therein). Moreover,
we find that the effective temperature of reddest FO models, at
fixed mass and luminosity, appears to be sensitive to the location
of the pulsation node within the convective stellar envelope,
deserving a detailed analysis that will be presented in a
forthcoming paper.

The lines drawn in Fig. 1 show, left to right, the location in the
HR diagram of the bluest FO, the bluest F and the reddest F
pulsators, respectively, for the labelled mass and varying the
luminosity under the assumption $l/H_p$=1.5. We find that, at
constant mass and luminosity, a variation of $Z$ from 0.0001 to
0.0004 has a marginal influence on the instability strip topology,
the only evident effect being the narrowing of the region where
only FO pulsation is efficient, that is the region between the
bluest FO and F models, when passing from $Z$=0.0001 to
$Z$=0.0004. As a whole, in spite of the different opacity tables,
the predicted topology of the instability strip is similar to the
one found by B97 for a stellar mass 1.5 $M_{\odot}$ and Z=0.0001
(see Fig. 1 in B97).

Inspection of the results plotted in Fig. 1 shows that  the entire
pulsation region tends to move towards the blue if increasing the
mass (at constant luminosity) or decreasing the luminosity (at
constant mass). A linear regression to the results yields that, in
the range $Z$=0.0001 to 0.0004, the limits of the AC instability
strip can be approximated as

$$\log T_e(FOBE)=3.947(\pm 0.006)-0.052\log L+0.042\log M\eqno(2)$$

$$\log T_e(FRE)=3.898(\pm 0.006)-0.065\log L+0.058\log M\eqno(3)$$

\noindent where the intrinsic uncertainty of $\pm$ 50 K on the
FOBE and FRE temperatures is included. Concerning the fundamental
blue edge (FBE), our models suggest log$T_e$(FBE)$-$log$T_e$(FRE)
$\sim$ 0.055$\pm$ 0.005, for fixed mass and luminosity. Note that,
as already discussed for RR Lyrae stars (see Caputo et al. 2000,
Marconi et al. 2003), predictions concerning the FOBE, that is the
boundary between non pulsating and pulsating stars at the hotter
effective temperatures, are particularly solid. In fact, the
uncertainties connected to the treatment of convection are
expected to be smaller in the blue region (high effective
temperatures) of the instability strip and specific tests show
that a variation of the mixing length parameter from $l/H_p$ from
1.5 (the adopted value in this investigation) to 2.0 implies that
the FOBE becomes redder by not more than 100 K whereas the FRE
becomes bluer by at least 200 K.

\subsection{Light curves and pulsation amplitudes}

For each investigated model, the variation of relevant parameters
(luminosity, radius, radial velocity, effective temperature and
gravity) along a pulsation cycle was provided. A sub-sample of
bolometric light curves for $Z$=0.0001 (with M/M$_\odot$=2.0 and
logL/L$_\odot$=2.08) is shown in Figs. 2a,b for fundamental and
first overtone pulsators, respectively. The behavior of the curves
shows that, at variance with the case of RR Lyrae stars, no clear
information on the pulsation mode can be drawn from the light
curve morphology. Indeed, even if the raising branch is steeper in
the case of the fundamental mode, the light curves for both modes
present a variety of morphological features, deviating from a
sinusoidal shape but for the few hottest first overtone pulsators.

Figure 3 shows bolometric amplitudes versus periods for selected
models with $Z$=0.0001. In particular, the effects of varying the
luminosity level at fixed stellar mass (2.2$M_{\odot}$, top panel)
and of varying the mass at fixed luminosity
(log$L/L_{\odot}$=2.08, bottom panel) are shown. As a whole,
fundamental pulsators follow the well known behavior of
fundamental RR Lyrae stars, with the amplitude increasing as the
period decreases, for fixed mass and luminosity. Also the
dependence on mass and luminosity is the same as found for RR
Lyrae stars (see Bono et al. 1997b, Marconi et al. 2003): at fixed
period, the amplitude increases with increasing the luminosity or
decreasing the mass. Moreover, the rather linear dependence of the
amplitude with period will allow us (see Section 3) to derive
fundamental Period-Amplitude relations in analogy with RR Lyrae
stars.

As for first overtone pulsators, let us recall that, for any fixed
mass and luminosity, a model with an effective temperature 100 K
higher than the hottest pulsating model is not pulsating. This
means that within $\delta$ log$P\sim $ 0.02-0.03 to the left of
the hottest (shortest period) FO model plotted in Fig. 3 the
amplitude should reach vanishing values. On this basis, the
amplitudes plotted in Fig. 3 appear to show the characteristic
{\it bell shape} typical of first overtone RR Lyrae stars.

However, in spite of these similarities, two main differences
appear to exist between the behaviors of  ACs and RR Lyrae stars
in the Bailey diagram:
\begin{enumerate}

\item in the case of ACs the  maximum amplitudes reached by first
overtone pulsators may be comparable with those of fundamental
ones, whereas fundamental RR Lyrae pulsators always attain
significantly higher amplitudes;

\item the maximum of the {\it bell} defined by first
overtone pulsators decreases when the stellar mass increases or
the luminosity level decreases, following an opposite trend with
respect to first overtone RR Lyrae stars (see Bono et al. 1997b
for details).

\end{enumerate}

Considering that observed ACs, even in the same stellar system,
may have different mass and luminosity, with the more massive ones
at brighter luminosities, one has that the discrimination of the
AC pulsation mode from the analysis of the period-amplitude
diagram might be not an easy task. As a fact, inspection of Fig. 3
reveals that the period-amplitude distribution of the more massive
(2.2$M_{\odot}$) and brighter (log$L/L_{\odot}$=2.28) FO pulsators
(dotted line connecting open circles in the top panel), is quite
close to that of less massive (1.6$M_{\odot}$) and fainter
(log$L/L_{\odot}$=2.08) F pulsators (solid line connecting dots in
the bottom panel).

\section{From theoretical to observational quantities}

In order to compare theoretical results with observations, the
bolometric light curves have to be transformed into the
photometric bands $UBVRIJK$. As in previous studies performed by
our group, we use bolometric corrections and temperature-color
relations provided by Castelli, Gratton \& Kurucz (1997a,b).

Before proceeding, it is worth noticing that for observed
pulsating stars we can only measure magnitudes and colors that are
time-averaged quantities over a pulsation period, rather than the
static values the stars would have were they not pulsating. For
this reason, in consideration that observed variables are fitted
with evolutionary tracks, it is important to estimate the mean
values from pulsation models and compare them with the
corresponding static quantities. A similar study has already been
performed for RR Lyrae stars (Bono, Caputo \& Stellingwerf 1995,
Marconi et al. 2003) and classical Cepheids (Caputo, Marconi \&
Ripepi 1999) and we are now in the position to extend the analysis
to ACs.

With $<M_i>$ and $(M_i)$ indicating intensity-weighted and
magnitude-weighted mean magnitudes, respectively, we show in Fig.
4 the difference between static $M_V$ and mean magnitudes as a
function of the pulsation visual amplitude $A_V$. One has that
intensity-weighted means are better representative of the static
values than magnitude-averaged quantities, in both cases the
discrepancy increasing toward higher amplitudes and less symmetric
light curves. As far as the $BV$ color is concerned, Fig. 5 shows
that increasing the visual amplitude the magnitude-averaged mean
colors $(B-V)$ tends to be redder than the corresponding static
values, whereas the opposite behavior holds for the
intensity-weighted mean colors $<B>-<V>$.

In general, the behaviors found for present models are in
agreement with the results already found for RR Lyrae stars and
classical Cepheids. In analogy with the procedure followed by
Marconi et al. (2003) for RR Lyrae stars, we derive the predicted
relation to estimate static $V_s$ magnitudes from observed
magnitude and intensity averaged values, as given by

$$V_s=1.20<V>-0.20(V)\eqno(4)$$

\noindent with an uncertainty of $\pm$ 0.01 mag. This relation
appears in substantial agreement with the results recently
presented for RR Lyrae models with $Z$=0.001 (see Marconi et al.
2003).

However, bearing in mind that the pulsation amplitude of models
close to the boundaries (FOBE and FRE) of the instability strip
approaches the zero value, the location of FOBE and FRE in the
$M_V$-log$P$ plane turns out to be quite independent of whether
static or mean magnitudes are considered. As shown in Fig. 6, a
linear regression to the models with $l/H_p$=1.5 yields the
analytical relations

$$M_V(FOBE)=-1.12-2.68\log P_{FO}-1.80\log M\eqno(5)$$

$$M_V(FRE)=-0.03-2.32\log P_F-1.92\log M\eqno(6)$$

\noindent with $\sigma_V$=$\pm$0.06 mag, including the intrinsic
$\pm$ 50 K uncertainty on FOBE and FRE effective temperatures.
Adopting $l/H_p$=2.0, the zero-point in eq. (5) and (6) changes
respectively by$\sim$ +0.1 mag and $\sim -$0.18 mag. As for the
blue limit of fundamental pulsation (FBE), where the amplitude
reaches quite large values, we consider intensity-weighted
magnitudes, deriving that the predicted location of the FBE can be
approximated as

$$<M_V(FBE)>=-0.72-2.61\log P_F-1.71\log M\eqno(7)$$

\noindent with $\sigma_V$=$\pm$0.07 mag.

\subsection{Period-Magnitude-Color, Color-Color relations and Wesenheit functions}

The natural outcome of the period relation (eqs. 1a, 1b) in the
observational plane is the Period-Magnitude-Color relation
(hereafter named PMC) in which the period is correlated with the
pulsator absolute magnitude and color, for any given mass. The
coefficients of the intensity-weighted PMC relations in the
various bands are reported in Table 2, while Fig. 7 shows the
quite small scatter of the models around the predicted relations.

According to these relations, one could estimate the mass range
spanned by a sample of ACs at the same distance and with the same
reddening with a formal uncertainty of $\sim$ 2\%, provided that
periods and colors are well measured. If the distance and
reddening are independently known, then the mass absolute values
can be determined too.

\begin{table}

\centering \caption[]{Coefficients of the mass-dependent
Period-Magnitude-Color relations.\label{tab2}}
\begin{tabular}{lccccc}

\hline
\\
Mode & $\alpha$  & $\beta$ & $\gamma$ & $\delta$ & $\sigma_V$  \\
\\
\hline \\

\multicolumn{6}{c}{$<M_V>=\alpha+\beta$log$P$+$\gamma[<B>-<V>]+\delta$log$M$}\\
F  & $-$1.56 & $-$2.85 & 3.51 & $-$1.88 & 0.01 \\
FO & $-$1.92 & $-$2.90 & 3.43 & $-$1.82 & 0.02 \\

\\

\multicolumn{6}{c}{$<M_V>=\alpha+\beta$log$P$+$\gamma[<V>-<R>]+\delta$log$M$}\\

F  & $-$1.69 & $-$2.90 & 5.39 & $-$1.89 & 0.01 \\
FO & $-$2.09 & $-$3.05 & 5.28 & $-$1.75 & 0.01 \\

\\
\multicolumn{6}{c}{$<M_V>=\alpha+\beta$log$P$+$\gamma[<V>-<I>]+\delta$log$M$}\\
F  & $-$1.83 & $-$2.93 & 2.73 & $-$1.89 & 0.01 \\
FO & $-$2.24 & $-$3.09 & 2.67 & $-$1.73 & 0.01 \\

\\

\multicolumn{6}{c}{$<M_V>=\alpha+\beta$log$P$+$\gamma[<V>-<J>]+\delta$log$M$}\\

F  & $-$1.95 & $-$3.01 & 1.78 & $-$1.87 & 0.01 \\
FO & $-$2.38 & $-$3.17 & 1.75 & $-$1.70 & 0.01 \\

\\

\multicolumn{6}{c}{$<M_V>=\alpha+\beta$log$P$+$\gamma[<V>-<K>]+\delta$log$M$}\\

F  & $-$1.99 & $-$3.05 & 1.31 & $-$1.86 & 0.01 \\
FO & $-$2.40 & $-$3.22 & 1.29 & $-$1.68 & 0.01 \\

\\

\multicolumn{6}{c}{$<M_V>=\alpha+\beta$log$P$+$\gamma[<J>-<K>]+\delta$log$M$}\\

F  & $-$1.97 & $-$3.07 & 2.17 & $-$1.85 & 0.01 \\
FO & $-$2.41 & $-$3.24 & 2.11 & $-$1.66 & 0.01 \\
\\
\hline
\end{tabular}
\end{table}
\begin{table}

\centering \caption[]{Coefficients of intensity averaged
Color-Color relations for fundamental models at $Z$=0.0001. The
last column gives the reddening vector, as estimated using a
typical extinction law (see text). \label{tab3}}

\begin{tabular}{lccccr}
\hline
\\
Mode & $a$  & $b$ & $\sigma$ & $r$  \\\\
\hline
\\

\multicolumn{6}{c}{$<B>-<V> = a + b [<V>-<R>]$}
\\
F  & -0.027 & 1.509 & 0.002 & 2.03 \\
FO & -0.016 & 1.461 & 0.004 & 2.03 \\

\\

\multicolumn{6}{c}{$<B>-<V> = a + b [<V>-<I>]$}
\\

F & -0.060  & 0.755 & 0.002 & 0.82\\
FO & -0.046 & 0.726 & 0.004 & 0.82\\

\\

\multicolumn{6}{c}{$<B>-<V> = a + b [<V>-<J>]$}

\\

F & -0.077  & 0.478 & 0.004 & 0.45 \\
FO & -0.062 & 0.461 & 0.007 & 0.45 \\

\\

\multicolumn{6}{c}{$<B>-<V> = a + b [<V>-<K>]$}

\\

F & -0.070  & 0.347 & 0.005 & 0.36\\
FO & -0.056 & 0.335 & 0.008 & 0.36\\

\\

\hline
\end{tabular}
\end{table}

We have also investigated the relations among intensity-averaged
colors, from optical to near infrared bands. As given in Table 3,
these relations have an intrinsic dispersion of the order of few
thousands of magnitude, but unfortunately they appear to be of no
use for what concerns the reddening determination. As a fact,
apart from the $<B>-<V>$  versus $<V>-<R>$ case, the slope $b$ of
the predicted relations is very close to the reddening vector $r$,
as estimated using a typical extinction law (see e.g. Cardelli,
Clayton \& Mathis 1989).

In order to complete the pulsational framework, we have finally
considered the reddening free Wasenheit functions (Madore 1982)
and their dependence on periods. As already discussed in several
papers on classical Cepheids (see, e.g., Madore \& Freedman 1991,
Tanvir 1999, Caputo, Marconi \& Musella 2000), the intrinsic
variation due to the finite width of the instability strip is
almost similar to the effects of interstellar extinction, and the
color term in the Wasenheit functions [$W(B,V)=M_V-3.1(B-V)$,
$W(V,I)=M_V-2.54(V-I)$, etc.], thought to cancel the reddening
effect, helps to reduce the dispersion of magnitudes at a given
period.  On this basis, the Period-Wesenheit relations (hereafter
referred to as PW) is used to get reliable estimates of the
intrinsic distance modulus.

This effect is shown in Fig. 8, while the coefficients of the
predicted intensity weighted PW relations, as derived by
fundamental models and first-overtone models with fundamentalised
periods, are reported in Table 4 for selected photometric bands.
Note that the use of these reddening-free relations to get the
mass range of a sample of ACs at the same distance yields a formal
uncertainty of $\sim$ 5\%, provided that periods and colors are
well measured. On the other hand, if the mass of the variable is
independently known, these relations can provide its intrinsic
distance modulus.

\begin{table}

\centering \caption[]{Coefficients of selected mass-dependent
fundamental PW relations $<W>=\alpha+\beta$log$P_F+\gamma$
log$M$.\label{tab4}}

\begin{tabular}{cccc}
\hline
\\\\
$\alpha$  & $\beta$ & $\gamma$ & $\sigma_V$  \\
\hline
\\
\multicolumn{4}{c}{$<W(B,V)>=<M_V>-3.10[<B>-<V>]$}\\
$-$1.43 & $-$2.77 & $-$1.83 & 0.04 \\

\\
\multicolumn{4}{c}{$<W(V,I)>=<M_V>-2.54[<V>-<I>]$}\\

$-$1.74 & $-$2.94 & $-$1.83 & 0.04 \\

\\

\multicolumn{4}{c}{$<W(V,K)>=<M_V>-1.13[<V>-<K>]$}\\

$-$1.73 & $-$2.93 & $-$1.83 & 0.04 \\
\\
\hline
\end{tabular}
\end{table}

\subsection{Period-Magnitude-Amplitude relation}

Once the bolometric light curves are transformed into the
observational plane, pulsation amplitudes in the various
photometric bands are derived. As already noticed in Sect. 2, the
behavior of fundamental pulsators in the Period-Amplitude diagram
is rather linear, but with a non negligible dependence on both the
pulsator mass and luminosity. This occurrence suggests the
possibility to derive a mass-dependent relation between period,
absolute magnitude and amplitude (PMA). A linear regression
through the fundamental models with $Z$=0.0001 and 0.0004 gives
the relation (see also Fig. 9)

$$\log P_F*=\log P_F+0.41<M_V>+0.77\log M=$$
$$0.01(\pm0.05)-0.188A_V\eqno(8)$$

\noindent that again should allow to estimate the mass range
spanned by fundamental ACs at the same distance and with well
measured periods and amplitudes, but with a formal uncertainty of
$\sim$ 15\% which is larger than that related with the PMC and PW
relations. However, it is of interest to note that, since the
PMA relation only holds for fundamental pulsators, a comparison
with the results inferred from the PMC or PW relations should
provide a way to discriminate the pulsation mode of observed ACs
(see below).

\section{Comparison with observations}

\subsection{Anomalous Cepheids in dwarf spheroidal galaxies}

As clearly shown by the evolutionary models (see Castellani \&
Degl'Innocenti 1995, B97, and the recent Pisa
(www.mporzio.astro.it/~marco/GIPSY/homegipsy.html) and Padova
(http://pleiadi.pd.astro.it) libraries), metal-poor central
He-burning structures more massive than $\sim$ 1.3$M_{\odot}$
evolve with a characteristic ``blue loop'' whose luminosity and
effective temperature increases when increasing the mass. Since
the effective temperature at the edges of the instability strip is
a function of mass and luminosity, it follows that one can easily
estimate whether a star with a given mass is expected to cross the
pulsation region becoming a variable star. Moreover, since at
fixed metal content the He-burning pulsators of a given mass would
have a quite defined range of intrinsic luminosity, the mass-term
in the above predicted relations could be removed to provide
``evolutionary'' relations.

However, from the point of view of stellar evolution  one should
consider that the mass/luminosity relation of central He-burning
structures is dependent on the metal content, as well as that,
even in presence of stellar models computed with the most updated
input physics, the He-burning luminosity of a given mass depends
on the adopted efficiency of core overshooting. As shown in Fig.
10 for two selected mass values, canonical He-burning models
computed with an inefficient overshooting (Pisa library) are
brighter  by $\sim$ 0.25 mag with respect to overshooting models
(Girardi et al. 2000). However, if for the given mass the
predicted FRE is taken into consideration, then the inclusion of
overshooting seems to leave almost unvaried the minimum mass for the
occurrence of AC variables. On this basis, in this paper we wish
to present only the results of pulsation models, leaving mass and
luminosity as free parameters and deserving to a forthcoming paper
the connection to stellar evolution theory and the construction of
synthetic relations based also on the evolutionary times.

In this section, we are going to apply the theoretical relations
presented in the previous section to  observed AC samples in dwarf
spheroidal galaxies with the purpose to estimate their mass. We
searched in the literature for  {\it bona fide} AC pulsators with
well measured periods, magnitudes and amplitudes and we found the
samples reported in Table 5 where the columns give the host
galaxy, the number of measured ACs, the available observed
quantities, the intrinsic distance modulus $\mu_0$, the $E(B-V)$
reddening (both $\mu_0$ and $E(B-V)$ as listed by Pritzl et al.
2002) and the corresponding references.

\begin{table*}

\centering \caption[]{Selected parameters of anomalous Cepheids in
dwarf spheroidal galaxies.}
\begin{tabular}{lcccccc}
\hline
\\
host  galaxy & AC number & observed quantities  & $\mu_0$ & $E(B-V)$ & source
\\\\
\hline

           &       & & & &        \\

And VI   & 6  & $V, B, A_V$ & 24.45 mag & 0.06 mag  & Pritzl et al. 2002\\
Leo II   & 4  & $V, A_V$    & 21.59 mag & 0.02 mag  & Pritzl et al. 2002\\
Draco    & 8  & $V, B, A_V$ & 19.49 mag & 0.03 mag  & Pritzl et al. 2002\\
Carina   & 8  & $V, B, A_V$ & 20.14 mag & 0.04 mag  & Dall'Ora et al. 2003  \\
Sculptor & 3  & $V, A_V$    & 19.56 mag & 0.02 mag  & Kaluzny et al. 1995\\
Sextans  & 6  & $V, B, A_V$ & 19.74 mag & 0.03 mag  & Mateo et al. 1995  \\
Fornax   & 17 & $V, I$      & 20.70 mag & 0.03 mag  & Bersier \& Wood 2002\\
\\
\hline
\end{tabular}
\end{table*}

A first comparison between theoretical results and observations is
presented in Fig. 11 where observed ACs are plotted in the
$M_V$-log$P$ diagram together with the predicted boundaries (FOBE
and FRE) of the instability strip at 1.3$M_{\odot}$ (dashed line)
and 2.2$M_{\odot}$ (solid line). We notice that the general
behavior of observed pulsators is in substantial agreement with
theory. Specifically, the comparison between the observed
distribution and the predicted edges confirms that the brightest
variables should have larger masses than the faintest ones.

For almost all the dwarf galaxies reported in Table 5, except
Carina and Fornax, a mode discrimination has been already
attempted (see Pritzl et al. 2002), as listed in the first column
of Table 6. Based on this classification and using the intrinsic
distance moduli and reddening values listed in Table 5, we proceed
in estimating the mass of individual variables from the predicted
PMC relations given in Table 2. The results are given in columns
(7) and (8) in Table 6. Furthermore, for  F pulsators only, the
mass is estimated also from the predicted PMA relation (see eq.
8), yielding the values listed in column (9) in Table 6. It seems
worth noticing that the stellar masses inferred in this paper
appear in substantial agreement with the mass range of ACs, as
predicted by evolutionary models of not-too-old metal-poor HB
structures(see Castellani \& Degl'Innocenti 1995, B97, and the
quoted Pisa and Padova libraries).

\begin{table*}

\centering \caption[]{Estimated mass values for F and FO anomalous
Cepheids, as inferred by PMC and PMA relations. The mode pulsation
given in the first column is taken by previous studies, while that
given in the last column is determined in the present paper (see
text). }

\begin{tabular}{lcccccccc}

\hline
\\
id & $P$ & $V$ & $B$ & $A_V$ & $M/M_{\odot}$ &

$M/M_{\odot}$ & $M/M_{\odot}$\\

&(days)&(mag) &(mag) &(mag) &(PMC$_F$) &(PMC$_{FO}$) &(PMA$_F$) &mode\\
\\
\hline
\\\\
And VI\\
V06-F   & 0.629 & 24.53 & 24.87 & 0.74 & 1.14 & --  &1.41   & F\\
V44-FO  & 0.760 & 23.62 & 23.99 & 0.53 &(2.95)& 1.84&(3.81) & FO\\
V52-FO  & 0.725 & 23.57 & 23.88 & 0.52 &(2.58)& 1.71&(4.30) & FO\\
V83-FO  & 0.674 & 23.47 & 23.79 & 0.50 &(3.46)& 2.21&(5.49) & FO\\
V84-F   & 1.357 & 23.66 & 24.05 & 0.60 & 1.29 & --    &1.63 & F\\
V93-F   & 0.477 & 24.75 & 25.13 & 0.60 & 1.57 & --    &1.67 & F \\
\\

Leo II  \\
V1-F    & 0.408 & 21.97 & --    & 0.76 & --    & --   &1.46 & --\\
V51-FO  & 0.396 & 21.59 & --    & 0.77 & --    & --   & --  & --\\
V27-F   & 1.486 & 20.45 & --    & 1.24 & --    & --   &1.34 & --\\
V203-F  & 1.380 & 20.59 & --    & 1.05 & --    & --   &1.38 & --\\
\\

Draco  \\
V55-F   & 0.552 & 19.44 & --    & 0.57 & --    & --   &1.93 & --\\
V119-F  & 0.907 & 19.03 & --    & 1.00 & --    & --   &1.31 & --\\
V134-FO & 0.592 & 18.78 & 19.06 & 0.88 & (2.58)& 1.66 &(3.35) &FO \\
V141-F  & 0.901 & 19.12 & 19.43 & 0.72 & 1.02  & --   &1.39 & F\\
V157-F  & 0.936 & 18.77 & 19.24 & 1.04 & 2.94  & --   &1.69 & F\\
V194-F  & 1.590 & 18.12 & 18.53 & 0.46 & 2.25  & --   &2.62 & F \\
V204-FO & 0.454 & 19.24 & 19.49 & 0.78 & (1.95)& 1.24 &(2.86) &FO\\
V208-F  & 0.608 & 19.28 & --    & 0.33 & --    & --   &2.37 & --\\
\\

Sextans\\
V9-FO   & 0.293 & 19.99 & 20.32 & 0.91 &(2.90) & 1.85 &(2.55) &FO\\
V34-FO  & 0.341 & 19.88 & 20.34 & --   & --    & 2.96 & --  & --\\
V5-F    & 0.862 & 19.51 & 19.83 & 0.68 & 0.94  & --   &1.26 & F\\
V19-FO  & 0.491 & 19.46 & 19.96 & --   & --    & 3.46 & --  & --\\
V6-F    & 0.926 & 19.18 & 19.47 & --   & 1.17  & --   & --  & --\\
V1-FO   & 0.693 & 18.83 & 19.08 & 0.82 & (2.28)& 1.41 &(3.61) &FO\\
\\

Sculptor \\
V5689-F  & 0.855 & 19.14 & --   & 0.70 & --    & --   &1.55 & --\\
V119-F   & 1.159 & 18.86 & --   & 0.55 & --    & --   &1.58 & --\\
V26-F    & 1.346 & 18.55 & --   & 0.80 & --    & --   &1.67 & --\\
\\

Carina \\
V14    & 0.475 & 19.84 & 20.13 & 0.70    & 2.21  & 1.43 &3.06 &FO\\
V29    & 0.718 & 18.97 & 19.25 & 0.82    & 3.31  & 2.15 &4.88 &FO\\
V33    & 0.590 & 19.76 & 20.10 & 0.63    & 2.25  & 1.44 &2.67 &FO\\
V158   & 0.630 & 20.33 & 20.73 & 0.46    & 1.32  & 0.82 &1.34 &F\\
V182   & 0.785 & 20.13 & 20.53 & 0.34    & 1.18  & 0.73 &1.37 &F\\
V187   & 0.475 & 19.21 & 19.41 & 1.10    & 3.22  & 2.13 &5.30 &FO\\
V193   & 0.424 & 19.26 & 19.50 & 0.15    & 4.39  & 2.94 &9.91 &FO\\
V205   & 0.383 & 19.63 & 19.85 & 0.16    & 3.03  & 2.01 &7.12 &FO\\
\\
\hline
\end{tabular}
\end{table*}

The mass values listed in Table 6 for the fundamental pulsators
are plotted in the upper panel in Fig. 12, together with the
formal errors as given by the intrinsic dispersion of the
predicted PMC and PMA relations. The same intrinsic dispersions
yield that the two mass values should agree at a level of
$\delta$log $M \sim \pm$0.07 (solid lines). Allowing for further
uncertainties as due to photometric errors and light curve
fitting, the results plotted in the upper panel in Fig. 12 confirm
that for all the fundamental candidates, except V157 in Draco (FO
pulsator?), the mass values predicted by the two relations agree
with each other within the errors, even though the mass suggested
by the PMC relation appears systematically smaller ($\Delta$log $M
\sim -$0.09) than the PMA results. This effect could be connected
to the theoretical uncertainty on the $l/H_p$ parameter which
affects pulsation amplitudes and in turn the PMA relation. This
also means that one could in principle calibrate this free
parameter by requiring the agreement between the PMA results  and
those inferred by the PMC, which is independent of the adopted
$l/H_p$ value. Finally, it is of interest to notice that the
difference between the two mass values is independent of errors on
the apparent distance modulus, whereas it depends on the adopted
reddening. Moreover, the mass absolute values listed in Table 6
should vary by about $\pm$ 12\% with $\delta \mu$=$\pm$ 0.1 mag.

Let us now assume that also the FO variables are actually
fundamental pulsators. Using again the fundamental PMC and PMA
relations, one derives the quite unrealistic large masses given in
brackets in Table 6. Moreover, as shown in the lower panel in Fig.
12, the difference between the two mass values is now
significantly larger ($\Delta$log$M\sim -$0.17) than the allowed
uncertainty, with the exception of V9 in Sextans. In summary,
observed data seems to support the hypothesis that, if the
variable is a fundamental pulsator, then the mass values inferred
from the predicted PMC and PMA relations are
 internally consistent within the errors.
However the location of V9 in
Sextans, if this variable is a FO pulsator,
 would suggest that this condition is necessary but not sufficient
to discriminate the fundamental mode. On this ground, according to
the mass values listed in Table 6, let us adopt that V158 and V182
in Carina are {\it bona fide} fundamental pulsators, whereas V29,
V187, V193 and V205 should pulsate in the first-overtone mode. As
for V33 and V14, no discrimination is presently made.

Once the pulsation mode is discriminated, F and FO variables in
Table 6 can be used to get some useful empirical relations. We
show in Fig. 13 that the Period-Magnitude (PM) relations for
fundamental (including V158 and V182 in Carina) and first-overtone
pulsators (including V29, V187 and V193 in Carina) can be linearly
approximated as

$$<M_V>_F=-0.69(\pm0.14)-2.73\log P_F,\eqno(9a)$$

$$<M_V>_{FO}=-1.56(\pm0.25)-2.95\log P_{FO},\eqno(9b)$$

\noindent which are in fair agreement with the relations recently
provided by Pritzl et a. (2002).

Moreover, using only the variables with $BV$ data, we plot in Fig.
14 the absolute $<WBV>$ quantities versus period. We derive that
the PW relations are

$$<WBV>_F=-1.72(\pm0.20)-3.34\log P_F,\eqno(10a)$$

$$<WBV>_{FO}=-2.26(\pm0.20)-2.44\log P_{FO}\eqno(10b)$$

Finally, using only fundamental variables, we show in Fig. 15 that
the empirical fundamental PMA relation can be written as

$$<M_V>_F=-0.37(\pm0.18)-2.57(\log P_F+0.188A_V)\eqno(11)$$

In these figures, the arrows depict the variables V157 in Draco
(filled triangle) and V9 in Sextans (open dot), while the small
crosses show V14 and V33 in Carina for which there is no clear
classification. By inspection of the three figures, one derives
that V157 in Draco follows the fundamental PM and PMA relations,
but the PW relation for FO pulsators, likely suggesting that its
measured $BV$ color is too red. As for V9 in Sextans and the two
variables in Carina, one derives that they follow more closely the
FO relations rather than the fundamental ones.

In conclusion, the mode discrimination for
a sample
of anomalous Cepheids requires the knowledge of
periods, colors and amplitudes.
Assuming that the variables
listed in Table 6 are representative of the average properties
of ACs,
fundamental or first-overtone candidates
should agree with all the above empirical relations.

\begin{table*}
\centering \caption[]{Estimated mass values anomalous Cepheids in
Fornax, as inferred by PMC relations. The mode pulsation given in
the last column is determined in the present paper (see text). }

\begin{tabular}{lccccccc}

\hline
\\
id &  $P$ & $V$ & $V-I$ & $M/M_{\odot}$ & $M/M_{\odot}$ & mode \\

   &  (days) &(mag) & (mag) & (PMC$_F$) & (PMC$_{FO})$ &         \\

 J023843.0  & 1.045 &  19.80 & 0.56 & 1.91 & 1.12 & (F-FO) \\
 J023852.4  & 1.250 &  19.91 & 0.60 & 1.47 & 0.83 & (F) \\
 J024017.4  & 1.198 &  19.99 & 0.61 & 1.44 & 0.81 & (F) \\
 J023952.5  & 1.311 &  20.25 & 0.90 & 2.44 & 1.40 & (FO)\\
 J024002.7  & 0.533 &  20.36 & 0.85 & 7.18 & 4.96 & ? \\
 J024022.8  & 0.838 &  20.48 & 0.53 & 1.05 & 0.60 & (F) \\
 J023946.2  & 0.922 &  20.57 & 0.79 & 1.94 & 1.13 & (F) \\
 J023941.5  & 0.573 &  20.62 & 0.48 & 1.36 & 0.82 & (F) \\
 J023937.7  & 0.546 &  20.65 & 0.73 & 3.24 & 2.10 & (F-FO) \\
 J024058.3  & 0.481 &  20.79 & 0.66 & 2.70 & 1.75 & (F-FO) \\
 J023907.1  & 0.508 &  20.79 & 0.77 & 3.60 & 2.35 & (FO) \\
 J024050.2  & 0.506 &  20.83 & 0.73 & 2.97 & 1.92 & (F-FO) \\
 J024000.9  & 0.416 &  20.83 & 0.48 & 1.74 & 1.10 & (F-FO) \\
 J023926.8  & 0.504 &  20.83 & 0.67 & 2.47 & 1.58 & (F-FO) \\
 J024016.9  & 0.533 &  20.84 & 0.43 & 1.00 & 0.59 & (F) \\
 J023953.5  & 0.611 &  20.85 & 0.66 & 1.70 & 1.03 & (F) \\
 J023954.8  & 0.574 &  21.02 & 0.48 & 0.83 & 0.48 & (RR?) \\
\hline
\end{tabular}
\end{table*}

In order to show the importance of both color and amplitude
measurements, let us take into consideration the ACs observed in
Fornax by Bersier \& Wood (2002). Using the $VI$ data listed in
Table 7, the predicted PMC relations for F and FO pulsators (see
Table 2)  yield the mass values listed in columns (5) and (6),
respectively. As a whole, the results are consistent with the mass
values listed in columns (6) and (7) in Table 6, except the
variable J024002.7 which appears to be too  massive. Excluding
this variable, we show in the upper panel of Fig. 16 the
comparison between observed magnitudes, corrected with a visual
distance modulus $\mu_V$=20.79 mag (from the distance and
reddening values reported in Table 5), and the above empirical PL
relations for F and FO variables. One has that most of the
variables appear to be fundamental pulsators.

On the other hand, since our pulsation models give

$$<WVI>=-0.22(\pm0.001)+1.03<WBV>\eqno(12)$$
\noindent eqs. (10a) and (10b) can be easily transformed into

$$<WVI>_F=-1.93(\pm0.20)-3.34\log P_F,\eqno(13a)$$

$$<WVI>_{FO}=-2.52(\pm0.20)-2.51\log P_{FO}\eqno(13b)$$

\noindent giving us a further opportunity for determining the
pulsation mode. As shown in the lower panel of Fig. 16, there are
now at least six variables which fit quite well the empirical FO
relation. Such a discrepancy may be due to poor light curves in
the $I$ band, but without the further constrains provided by the
PMA relation, almost for half of the whole sample of variables in
Fornax we are unable to clearly discriminate between F and FO
pulsators.

\subsection{The case of Leo A}

Recent results of a search for short-period variable stars in the
metal-poor ($Z\sim$ 0.0004) Leo A dwarf irregular galaxy are
presented by Dolphin et al. (2002). These authors have found 66
variables brighter than RR Lyrae stars, noticing that there is
some ambiguity as to whether these objects should be classified.
Eventually, they conclude that these variables could reproduce the
natural extension to low metal abundance of classical Cepheids,
rather than being ``anomalous" Cepheids.

We show in Fig. 17 the $M_V$-log$P$ diagram of variables with
unambiguous period and quality rating $\ge$ 3, adopting an
apparent distance modulus of 24.6 mag (see Dolphin et al. 2002).
Filled and open symbols refer to F and FO pulsators, respectively,
as given in Dolphin et al. (2002). The comparison with the data in
Fig. 11 shows that the ``faint" ($M_V\ge -$0.5 mag) variables
observed in dwarf spheroidal galaxies are absent in Leo A. This
evidence would suggest that the variables in Leo A are on average
more massive than ACs in dwarf spheroidal galaxies.

As a fact, the comparison with the predicted boundaries of the
instability strip with $M$=2.2$M_{\odot}$ (solid lines), discloses
a quite reasonable agreement at the low luminosities, but the occurrence of
bright ($M_V \sim -$1 to $-$1.5 mag) first overtone candidates
located to the left of the predicted FOBE indicates the presence
of even larger masses. In order to verify such a hypothesis, we
have computed few models with 4.0$M_{\odot}$, finding that the
predicted FOBE at such a mass (dotted line) fits quite well  the
observed distribution. Based on such a result, we believe that the
variables in Leo A are a mix of ``anomalous" and ``classical"
Cepheids. A critical analysis of these variables requires both
pulsational and evolutionary models, and this issue will be
addressed in a forthcoming paper (Caputo et al. in preparation).

\section{Concluding remarks}

Based on nonlinear, nonlocal and time-dependent convective
pulsating models, we have presented a pulsational scenario for the
analysis of ``anomalous'' Cepheid (AC) stars, showing that a
variation of metallicity from $Z$=0.0001 to 0.0004 has a very mild
effect on the edges of the instability strip, as well as on the
predicted relations connecting the pulsational (period, amplitude)
to the intrinsic stellar parameters (mass, luminosity, effective
temperature). The main results of the present study can be
summarized a follows.

\begin{enumerate}

\item   We derive the pulsation relation connecting period with the star
structural parameters $M,L,T_e$.

\item   We give the predicted relations of the boundaries of the
pulsation region in the log$L$-log$T_e$ and $M_V$-log$P$ planes,
for fixed mass. These relations, together with evolutionary models
computed under different physical conditions (e.g., with or
without overshooting) are the basic ingredients to evaluate
whether a star is expected to evolve into the instability strip
during its central He-burning phase.

\item   We show that the comparison between the mean magnitudes and
colors, as based on theoretical light curves transformed into the
observational bands, and the corresponding static values, follows
the general trend also shown by other classes of pulsating stars
(RR Lyrae, classical Cepheids).

\item
We derive the predicted Period-Magnitude-Color (PMC) relations,
showing that it can be used to estimate, with a formal uncertainty
of $\sim$ 2\%, the mass range spanned by a sample of well-measured
variables at the same distance and with the same reddening. If the
distance and reddening are independently known, then the mass
absolute values can be determined too.

\item   As usually adopted for classical
Cepheids, an important tool to derive absolute distances is given
by the theoretical reddening-free Wesenheit functions. These
relations (PW) can also be used to get the mass range of a sample
of ACs at the same distance with a formal accuracy of $\sim$ 5 \%.

\item   As already known for RR Lyrae stars, the amplitude of
fundamental pulsators is linearly dependent on the period
(logarithmic scale), for fixed mass and luminosity, providing a
mass-dependent Period-Magnitude-Amplitude (PMA) relation which
again can be adopted to estimate the mass range spanned by
fundamental variables at the same distance, but with a larger
formal uncertainty ($\sim$ 15\%) with respect to the other
methods.

\item   Since the PMA relation only holds for fundamental
pulsators, whereas PMC and PW relations can be applied both to the
fundamental and the first overtone mode, the comparison between
the masses evaluated by adopting the different kinds of relations
may help to attempt a discrimination of the pulsation mode.
Moreover, we show that the difference between the mass values
inferred from PMC and PMA relations is independent of errors on
the distance modulus, whereas the mass absolute values should vary
by about $\pm$ 12\% with $\delta \mu$=$\pm$ 0.1 mag.

\end{enumerate}

The comparison between the predicted edges of the instability
strip and the observed distribution of ACs belonging to a number of
dwarf spheroidal galaxies seems to support the predictive
capabilities of pulsation models. The only exception is found for
few first overtone variables in the irregular galaxy Leo A, that
fall to the left of the predicted blue boundary reproducing the
other variables, suggesting a mass of the order of 4 $M_{\odot}$.

By considering dwarf spheroidal galaxies for which a mode
discrimination of ACs is available in the literature and by
applying the predicted PMC and PMA relations to these variables,
we are able to conclude that if  the variable is a fundamental
pulsator, then both PMC and PMA yield mass values which are
internally consistent within the errors. On this basis, we find
that only two variables in Carina are {\it bona fide} fundamental
pulsators, whereas the remaining variables should pulsate in the
first-overtone mode.

On the basis of the quoted mode discrimination empirical PL and PW
relations  for fundamental and first overtone ACs in
dwarf galaxies, as well as the PA relation for fundamental
variables, are derived.

{\bf Finally, we notice that, as remarked by our referee, the intrinsic errors
of the results listed above are valid in the context of the adopted 
treatment for the convective transfer. Indeed, even if our formalism 
has the advantages of being nonlocal and time dependent, nevertheless
 it has turned out recently that the inclusion of additional equations in
the Reynolds momentum equations (see e.g. Canuto \& Dubovikov 1998)
improves the treatment of turbulence.}

\begin{acknowledgements}

We thank our anonymous referee for his/her helpful comments and corrections
that have considerably  improved the manuscript. We also thank G. Bono and
 V. Castellani for useful suggestions and
remarks. Model computations made use of resources granted by the
``Consorzio di Ricerca del Gran Sasso", according to Project 6
``Calcolo Evoluto e sue Applicazioni (RSV6) - Cluster C11/B".
This work was financially supported by MIUR/Cofin 2002, under the 
project ``Stellar Populations in Local Group Galaxies'' (Monica Tosi
coordinator) and MIUR/Cofin 2001, under the project ``Origin and evolution
of stellar populations in the Galactic Spheroid (Raffaele Gratton coordinator).

\end{acknowledgements}

\clearpage
\begin{figure}
\caption{Dependence of the FOBE (dashed line), FBE (dotted line)
and FRE (solid line) on mass and luminosity, for the two labelled
metal abundances.}
\end{figure}

\begin{figure}
\caption{Theoretical bolometric light curves for a subsample
(M=2M$_{\odot}$ log$L/L_{\odot}$=2.08) of fundamental (a) and
first overtone (b) models with $Z$=0.0001.}
\end{figure}

\begin{figure}
\caption{Bolometric amplitudes versus periods for selected models
with $Z$=0.0001 at varying the luminosity level at fixed stellar
mass (2.2$M_{\odot}$, top panel) and  the mass at fixed luminosity
(log$L/L_{\odot}$=2.08, bottom panel).}
\end{figure}

\begin{figure}
\caption{Difference between static $M_V$ and either intensity
weighted (top panels) or magnitude averaged (bottom panels) mean
magnitudes as a function of the pulsation amplitude for both
fundamental (F) and first overtone (FO) models.}
\end{figure}

\begin{figure}
\caption{The same but for the B-V color.}
\end{figure}

\begin{figure}
\caption{Limits of the pulsation region (FOBE and FRE) and blue
limit for fundamental pulsation (FBE) in the $M_V$-log$P$ plane.
The periods are normalized to the same mass (1.9$M_{\odot}$),
according to the predicted dependence (see text).}
\end{figure}

\begin{figure}
\caption{PMC relations in the labelled photometric bands for both
fundamental (filled circles) and first overtone(open circles)
models. The solid line represents the linear regression through
the data (see Table 2).}
\end{figure}

\begin{figure}
\caption{PW relations in the labelled photometric bands for
fundamental (filled circles) and first overtone(open circles)
models with fundamentalised periods. The solid line represents the
linear regression through the data (see Table 4).}
\end{figure}

\begin{figure}
\caption{PMA relations for fundamental models. The solid line
represents the linear regression through the data (see eq. 8). }
\end{figure}

\begin{figure}
\caption{Central He-burning evolutionary tracks for the labelled
masses and metal content as computed without (solid line) or with
overshooting (dashed line). The predicted FRE is also shown.}
\end{figure}

\begin{figure}
\caption{Comparison between the location of observed ACs in the
$M_V$-log$P$ diagram and the predicted boundaries of the
instability strip at 1.3$M_{\odot}$ (dashed line) and
2.2$M_{\odot}$ (solid line).}
\end{figure}

\begin{figure}
\caption{(upper panel) - Comparison between the mass values of
observed fundamental ACs (symbols as in Fig. 11), as inferred by
the predicted PMA and PMC relations. The solid lines depict the
allowed uncertainty. (lower panel) - The same, but for observed
first-overtone ACs treated like fundamental pulsators.}
\end{figure}

\begin{figure}
\caption{Empirical $M_V$-log$P$ relations for observed fundamental
(upper panel) and first-overtone ACs (lower panel).  The small
arrows show V157 in Draco and V9 in Sextans discussed in the text.
The solid lines are from eqs. (9a)-(9b) in this paper.}
\end{figure}

\begin{figure}
\caption{As in Fig. 13, but with empirical $<WBV>$ quantities. The
solid lines are from eqs. (10a)-(10b) in this paper. }
\end{figure}

\begin{figure}
\caption{Empirical PMA relation for fundamental AC. The small
arrows show V157 in Draco and V9 in Sextans and the solid line is
from eq. (11) in this paper.}
\end{figure}

\begin{figure}
\caption{(upper panel) - $M_V$ versus log$P$ for ACs in Fornax.
The solid lines are the same drawn in Fig. 13. (lower panel) -
$<WVI>$ versus log$P$ for the same variables. The solid lines are
from eqs. (10a)-(10b) in this paper. }
\end{figure}

\begin{figure}
\caption{$M_V$ versus log$P$ for F (filled dots) and FO (open
dots) variables in Leo A with unambiguous period and quality
rating $\ge$ 3, adopting an apparent distance modulus of 24.6 mag
(see Dolphin et al. 2002). The solid lines represent the predicted
boundaries of the instability strip with $M$=2.2$M_{\odot}$, while
the dotted line is the predicted FOBE at  $M$=4.0$M_{\odot}$(see
text for details).}
\end{figure}

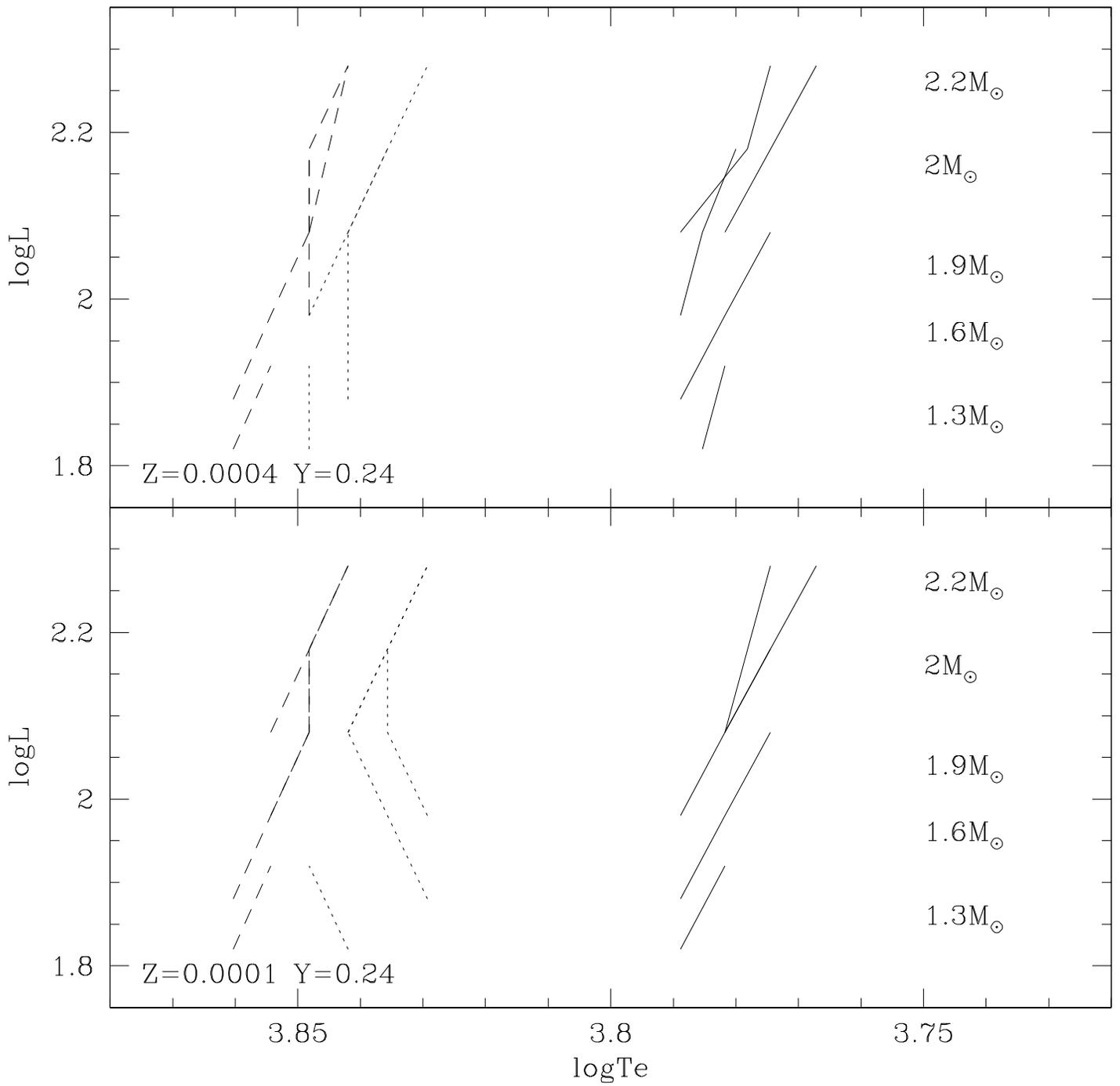
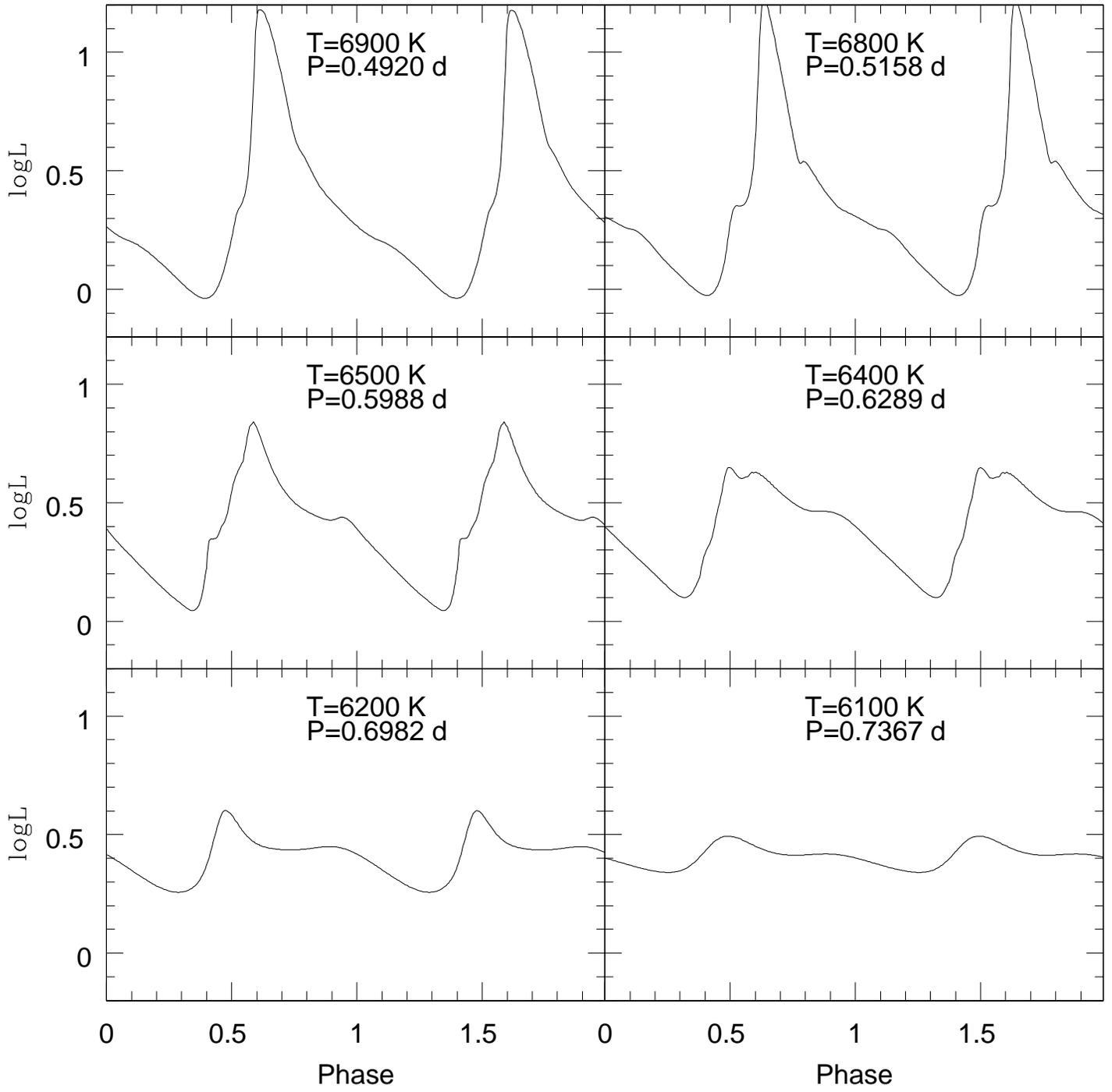
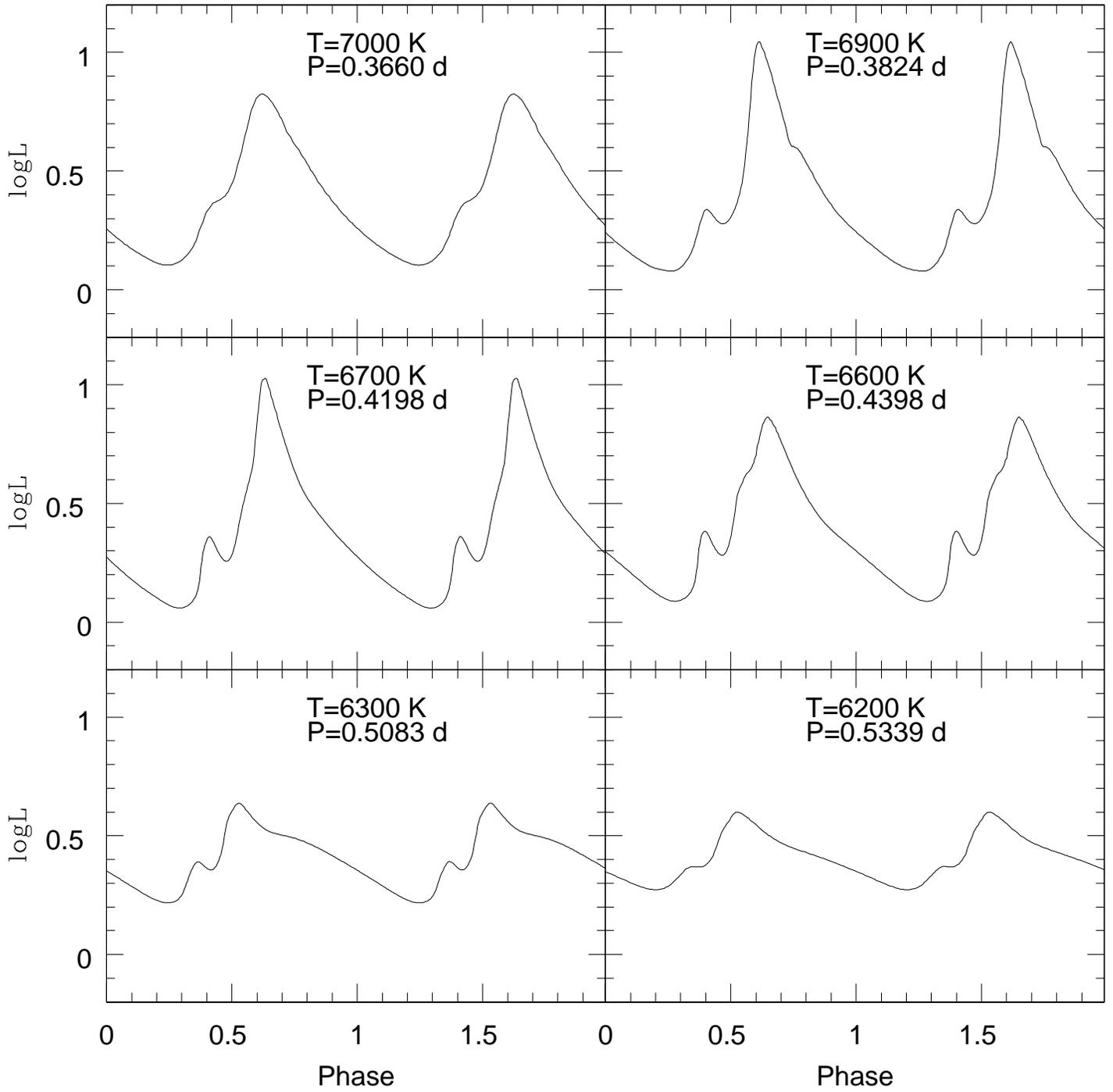
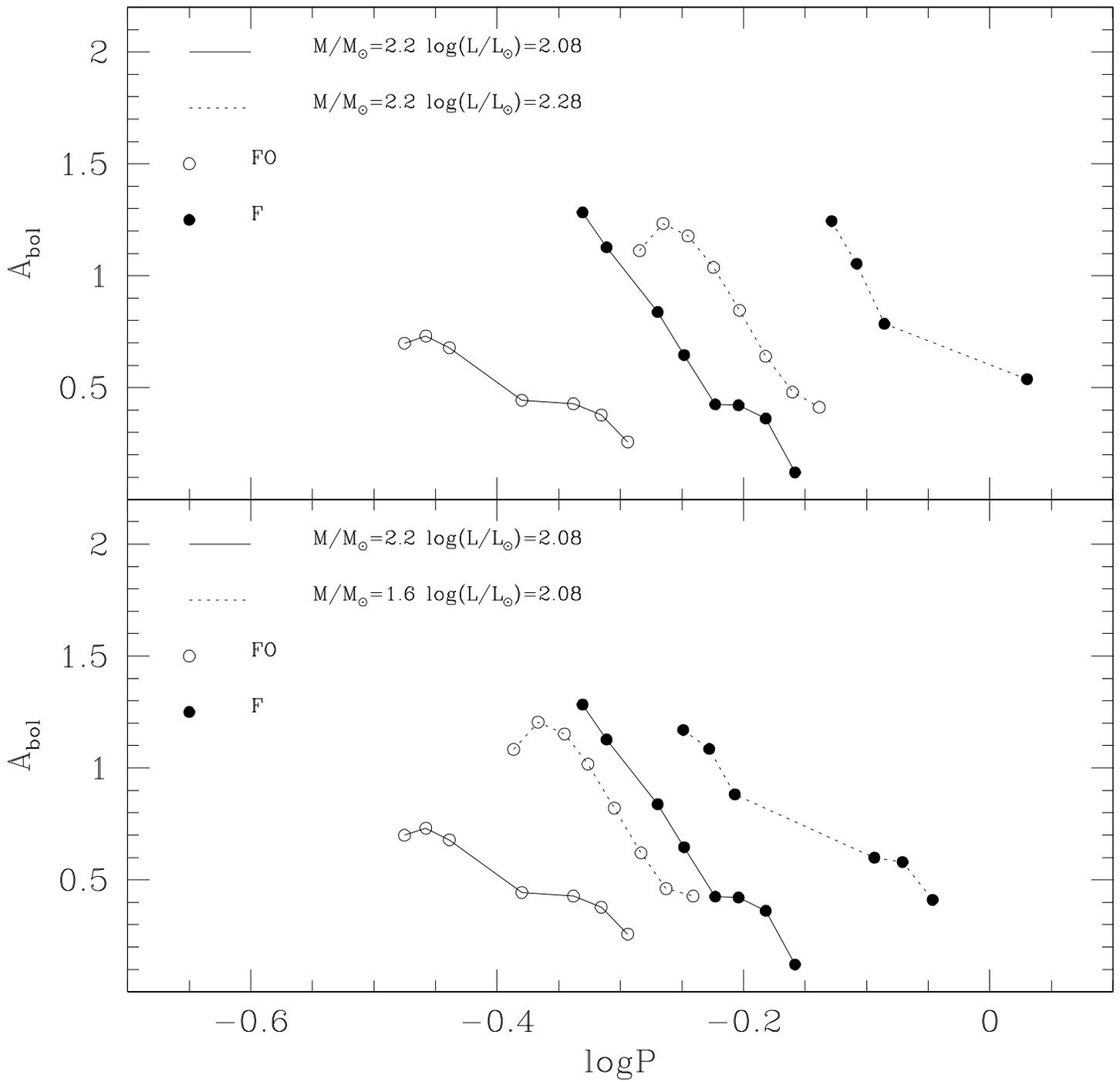
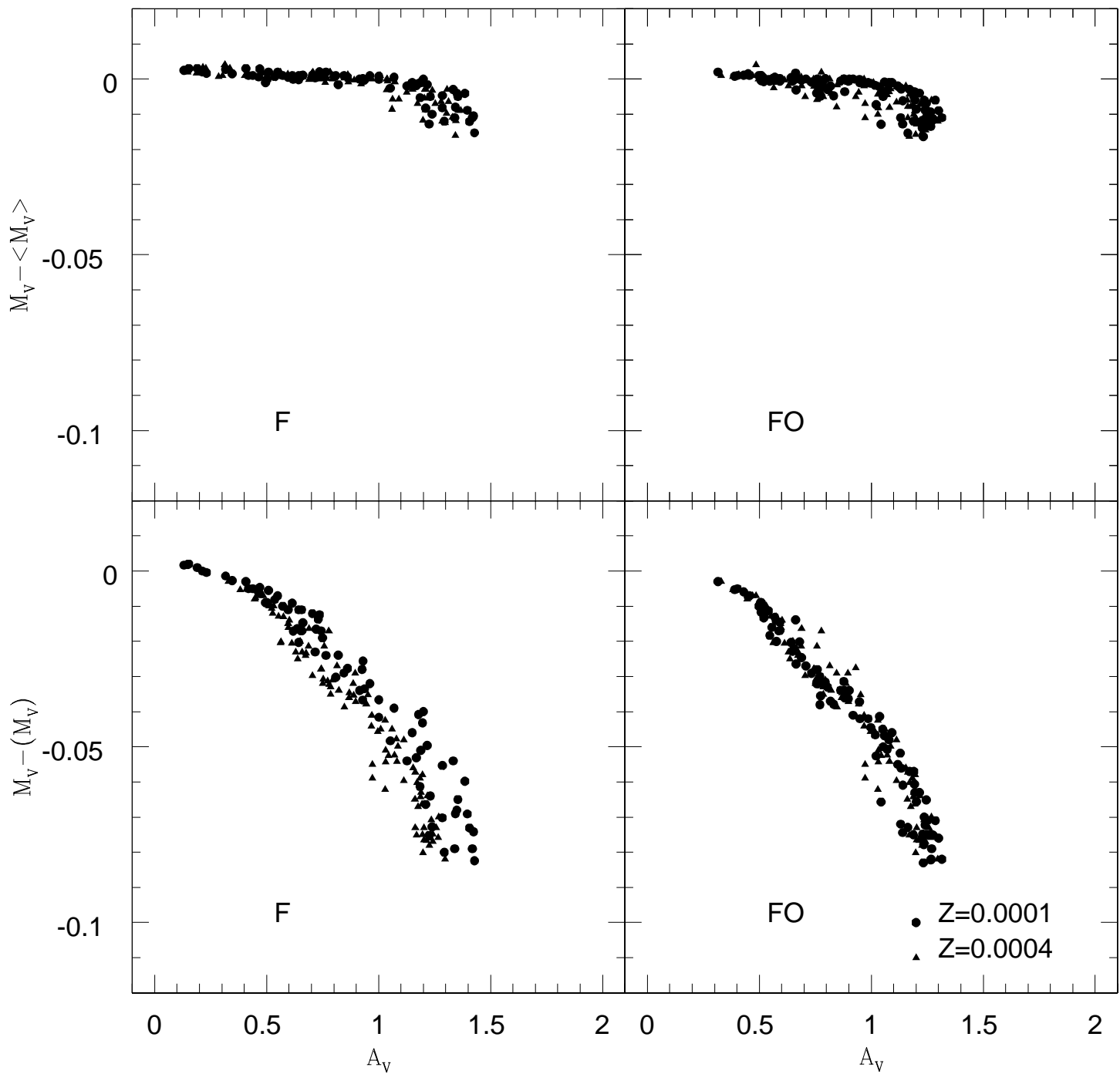
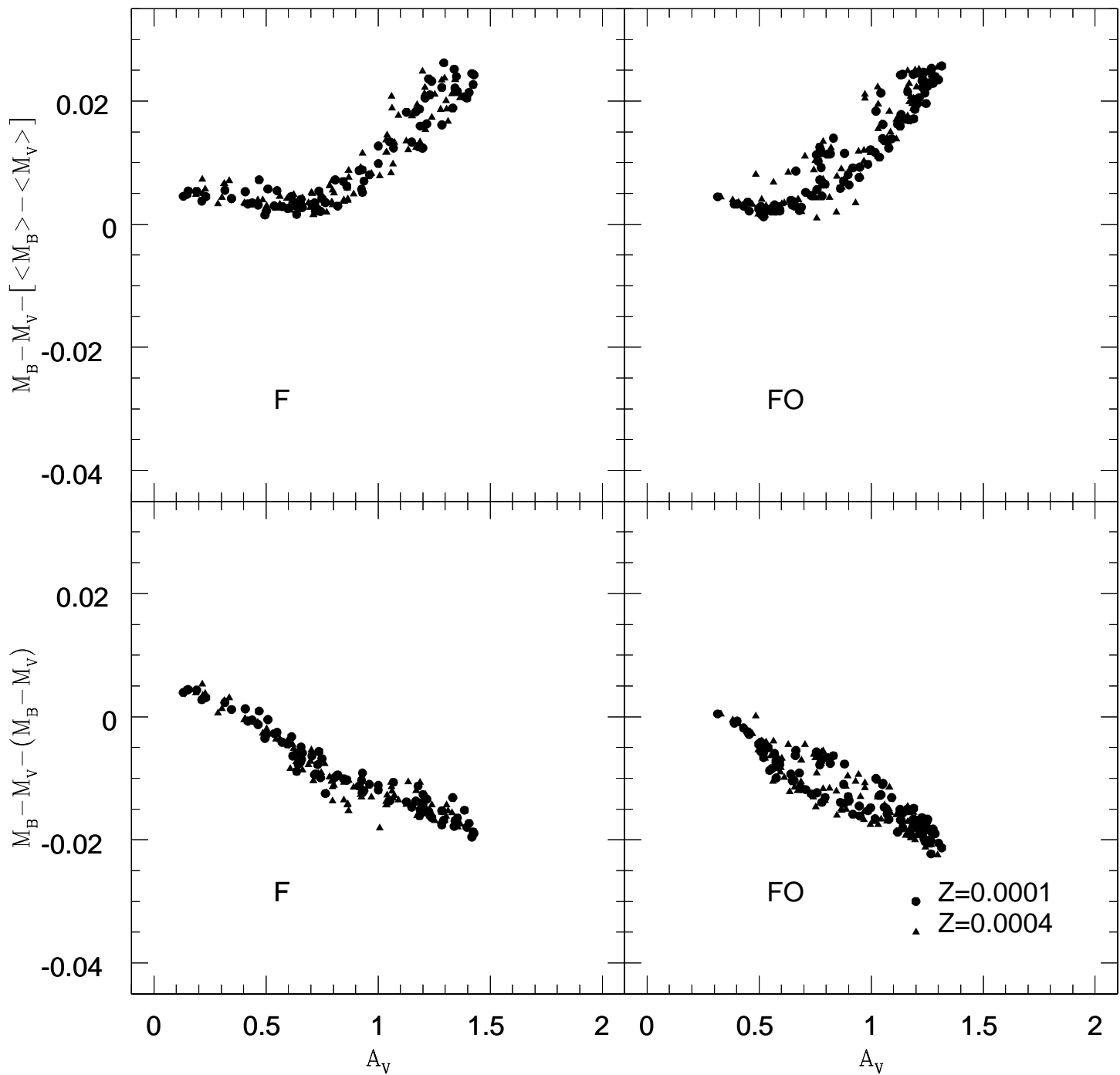
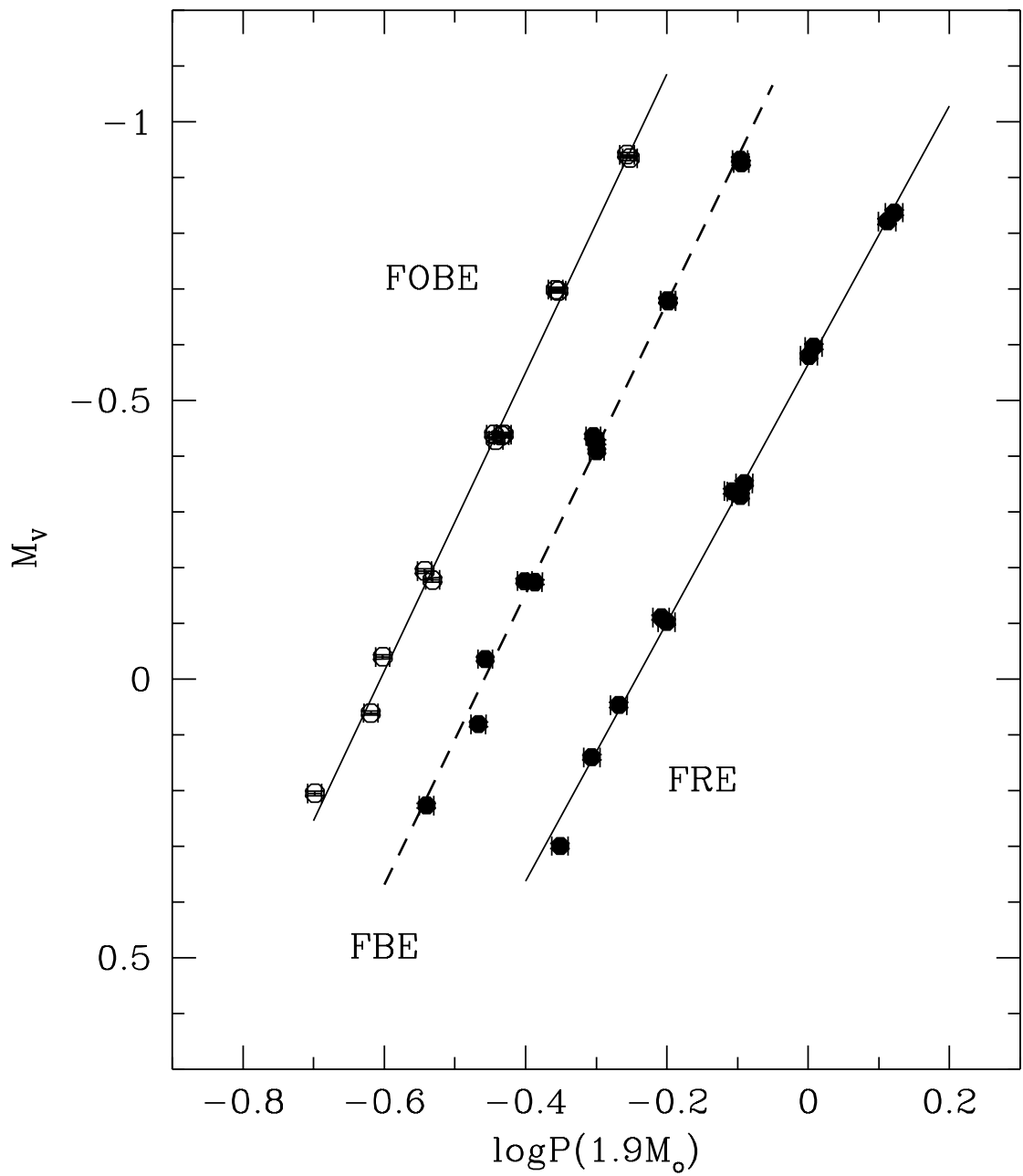
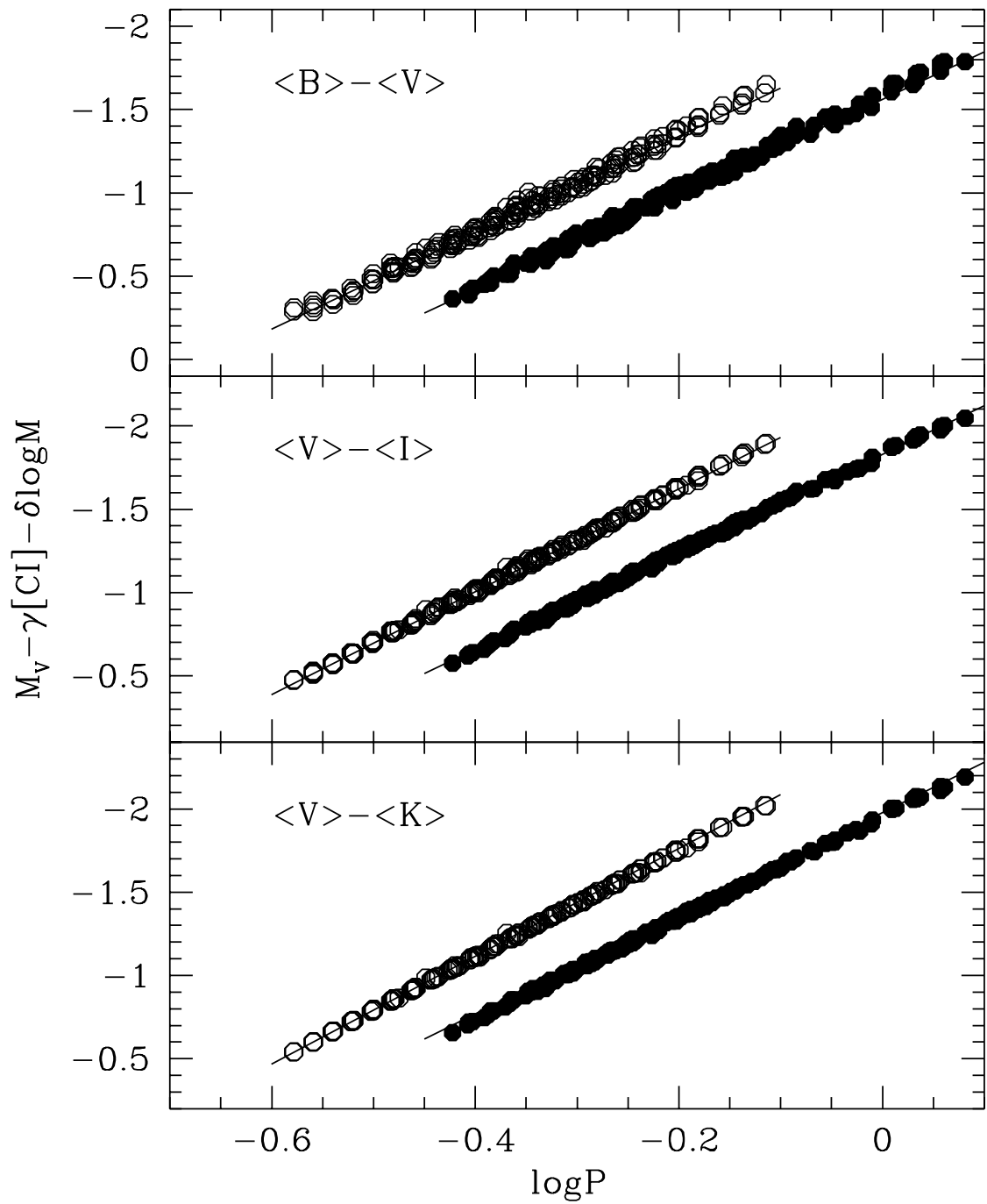
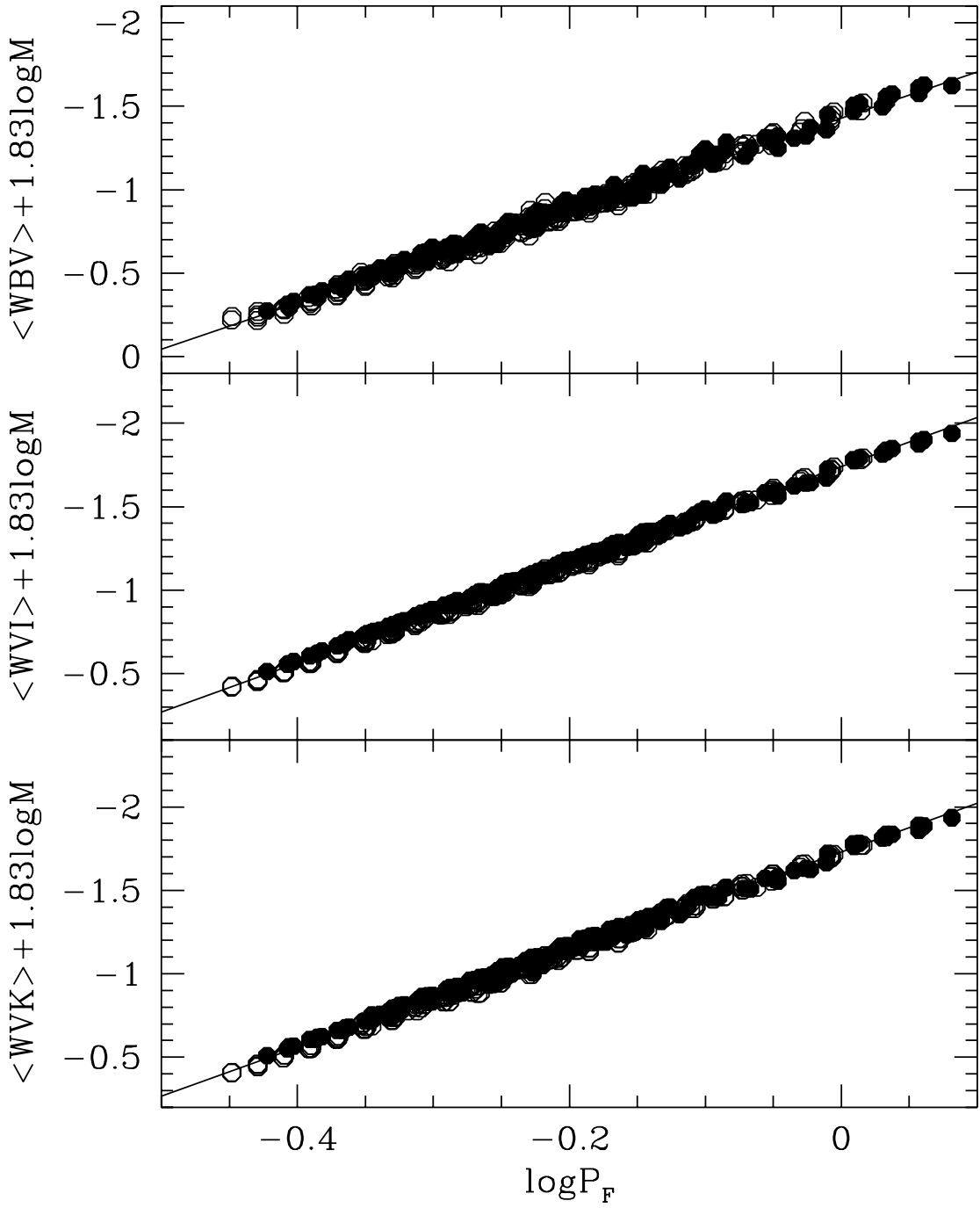
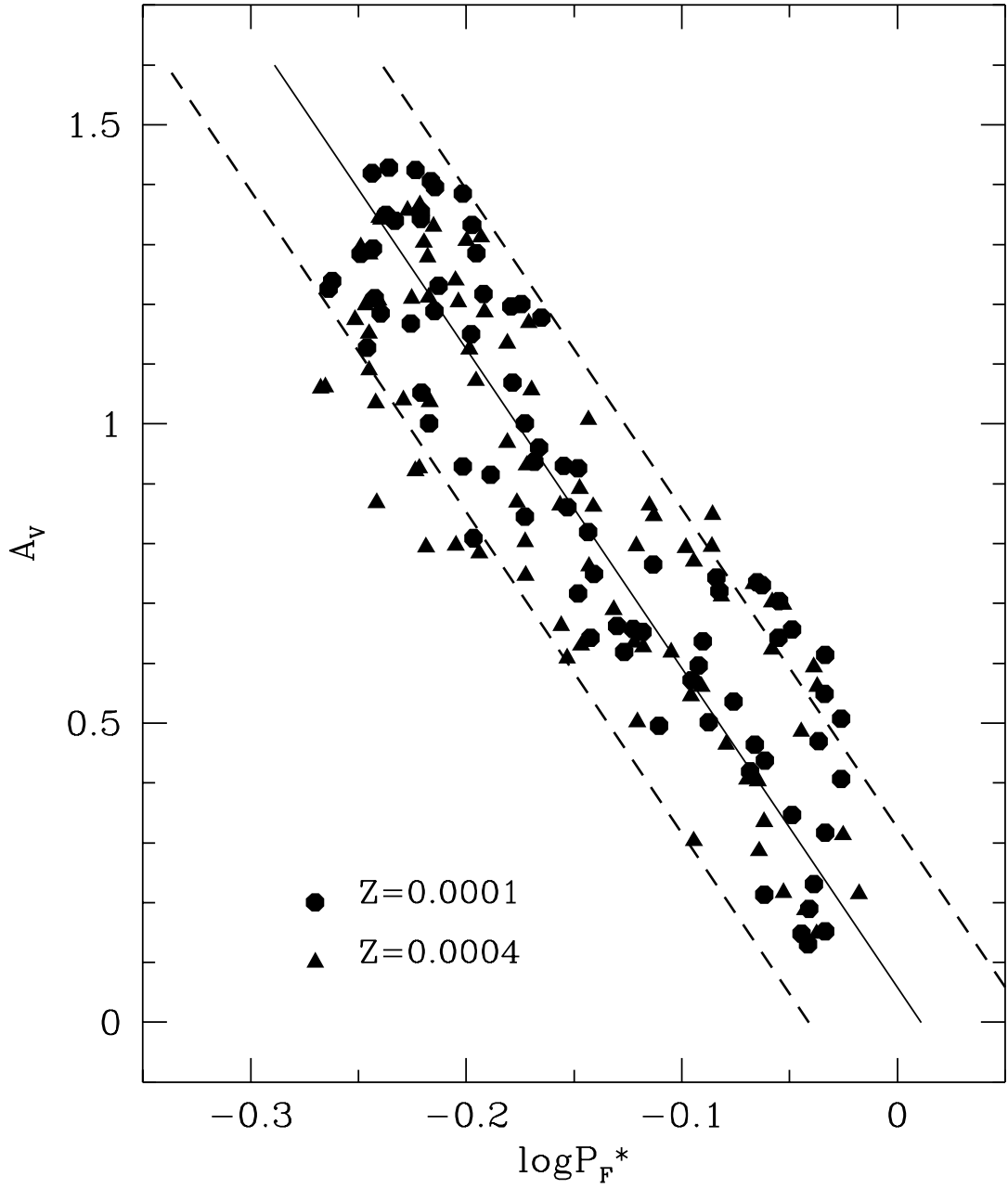
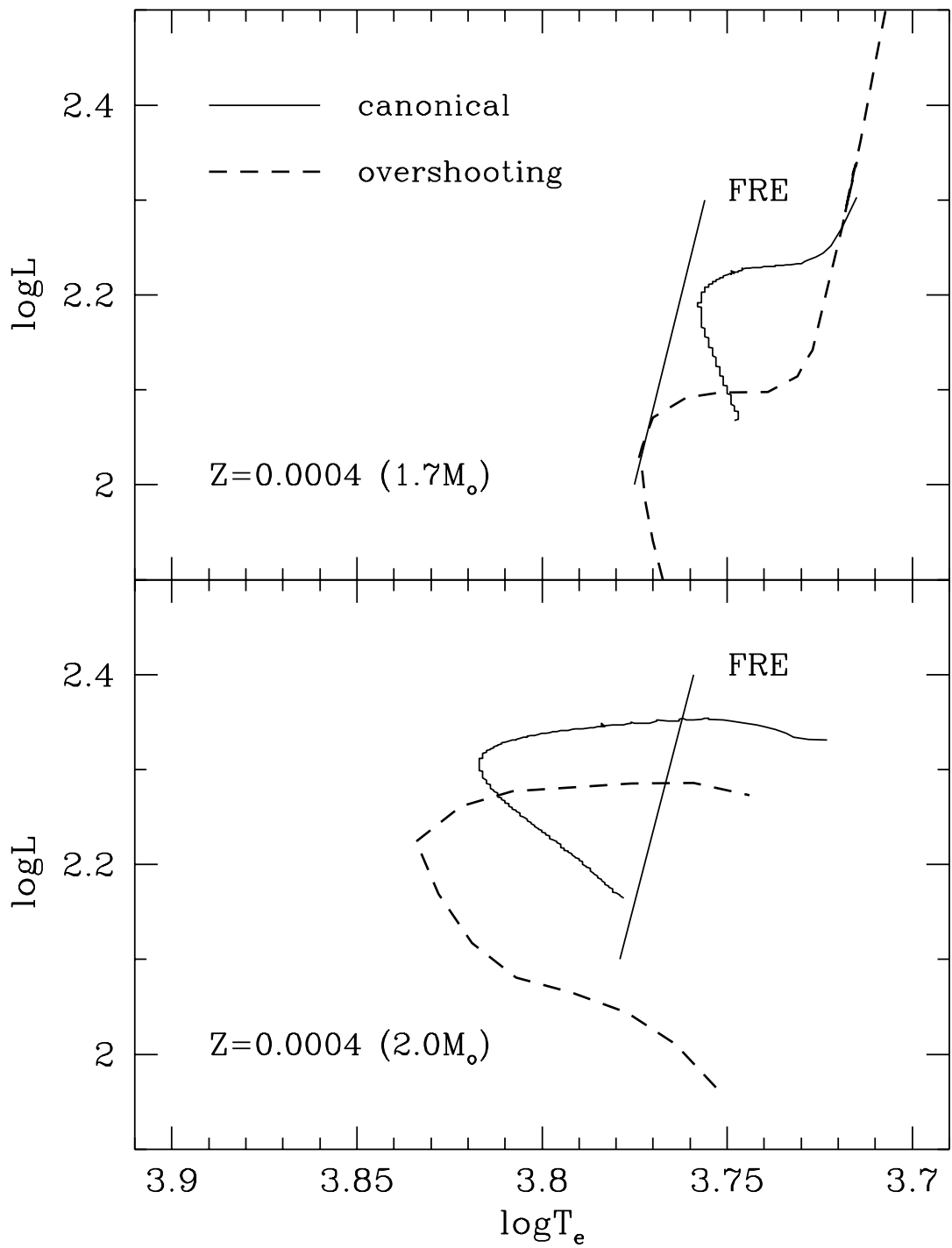
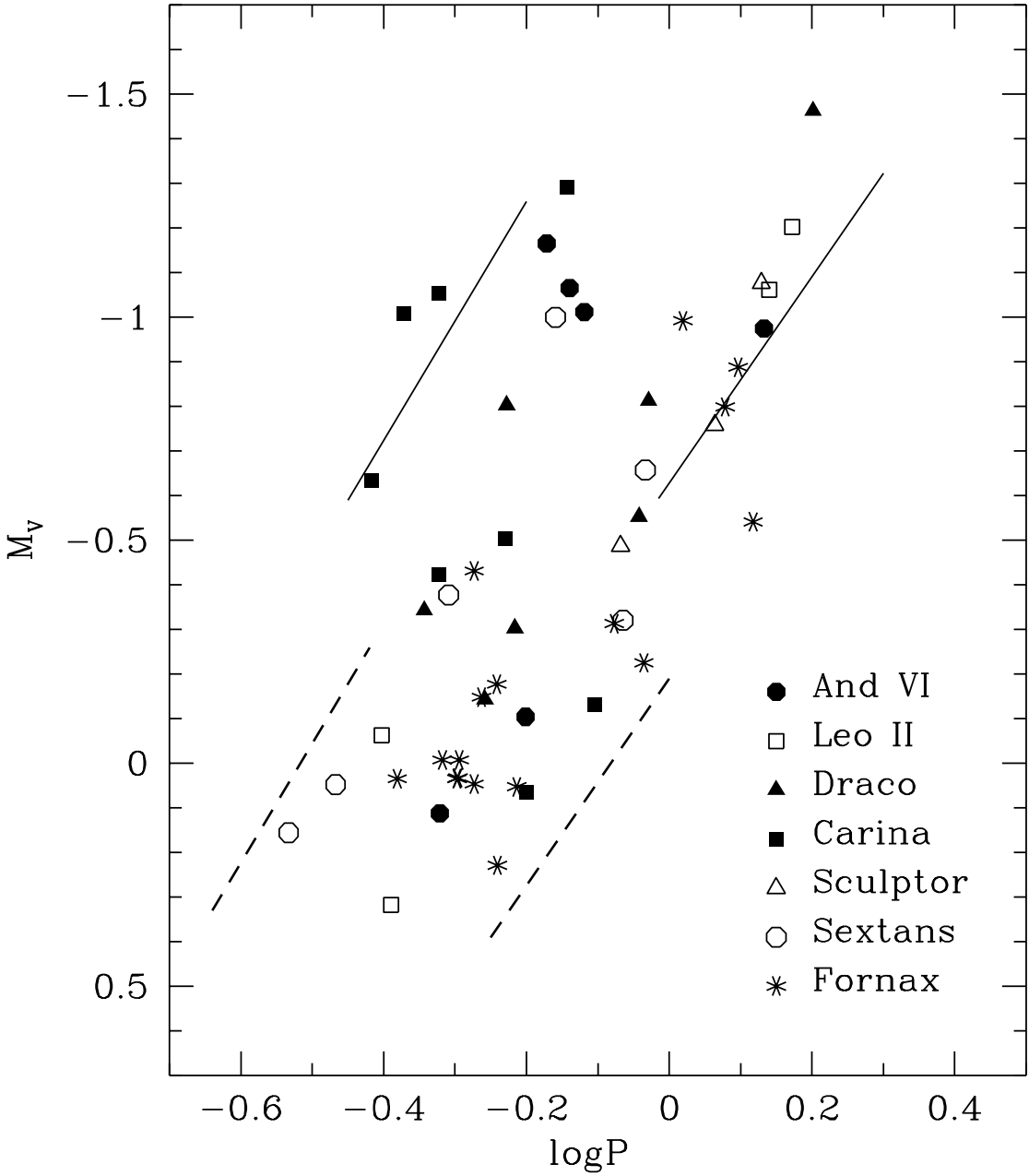
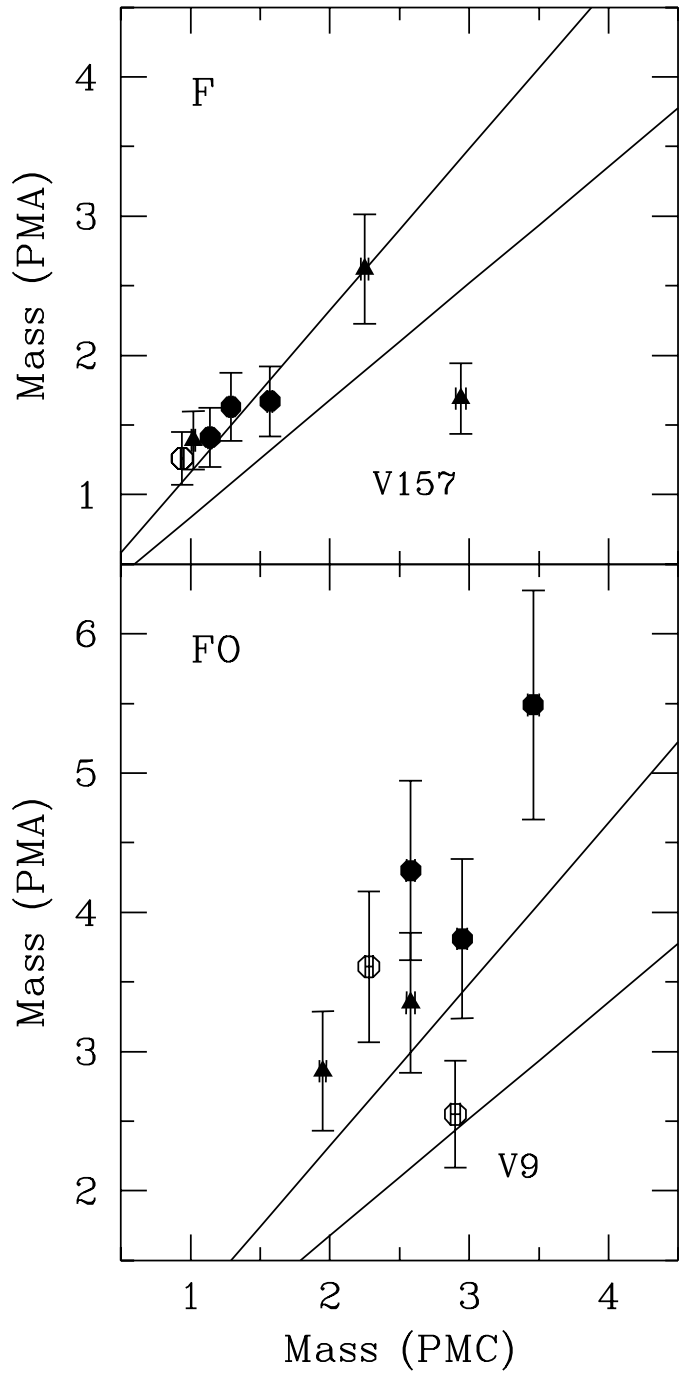
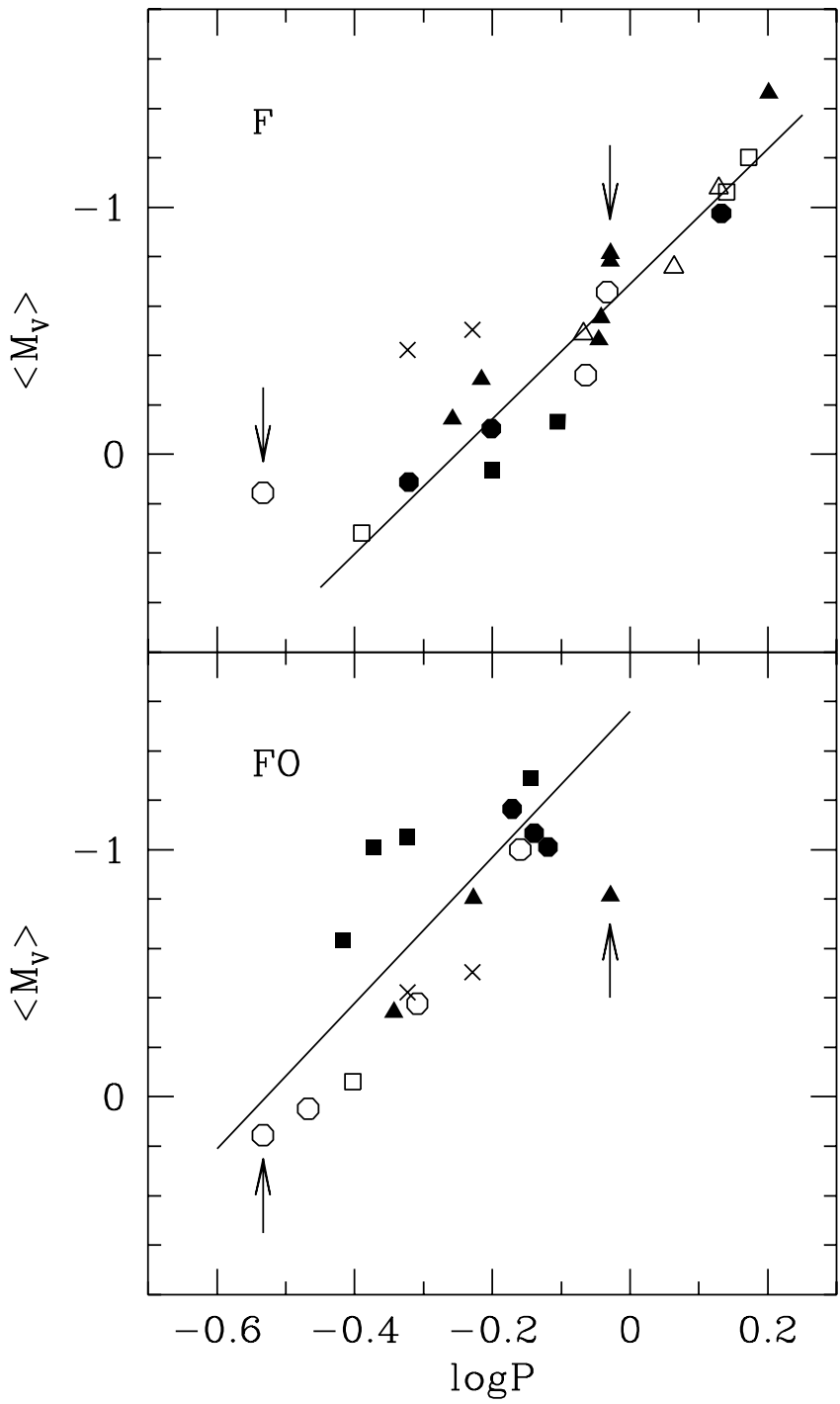
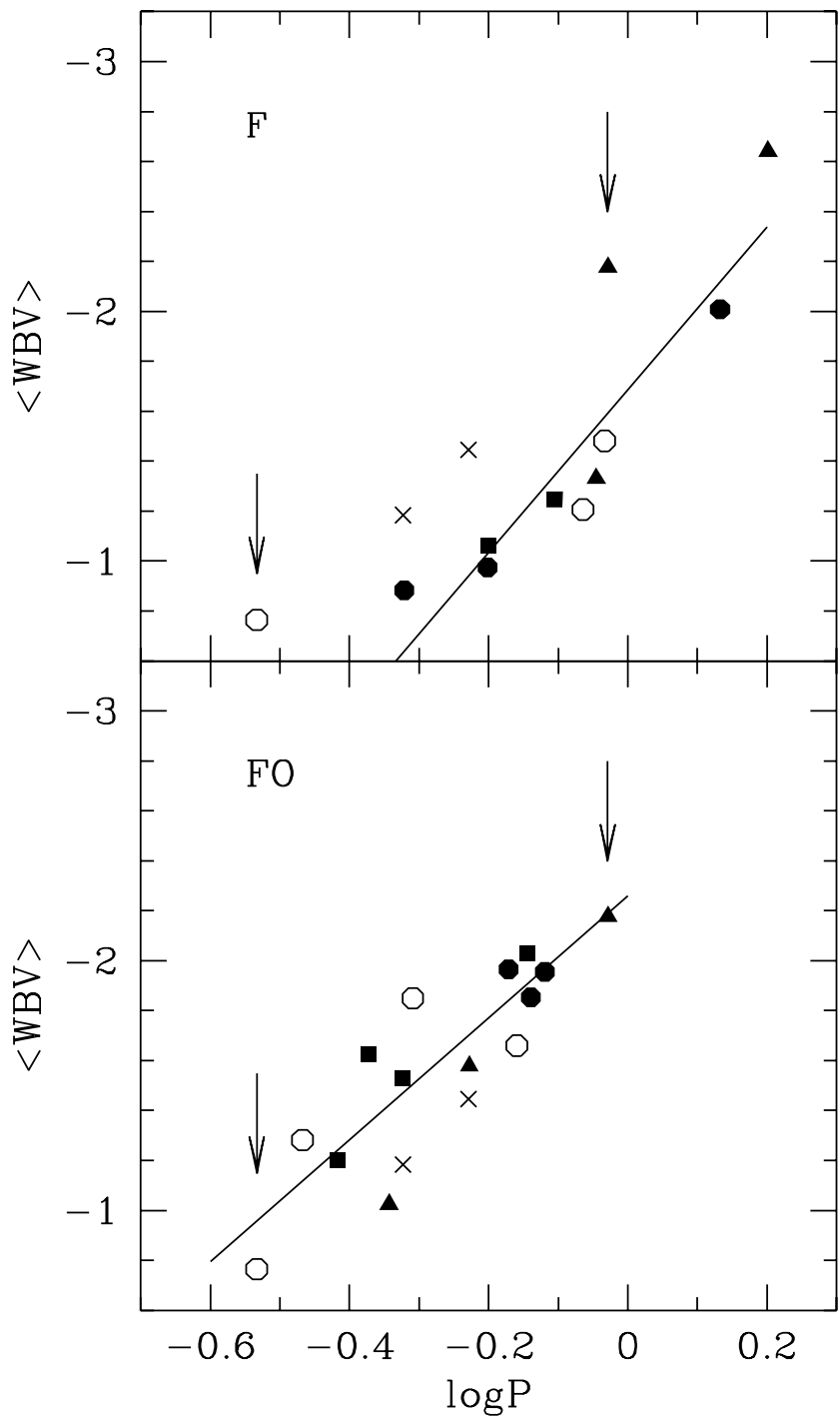
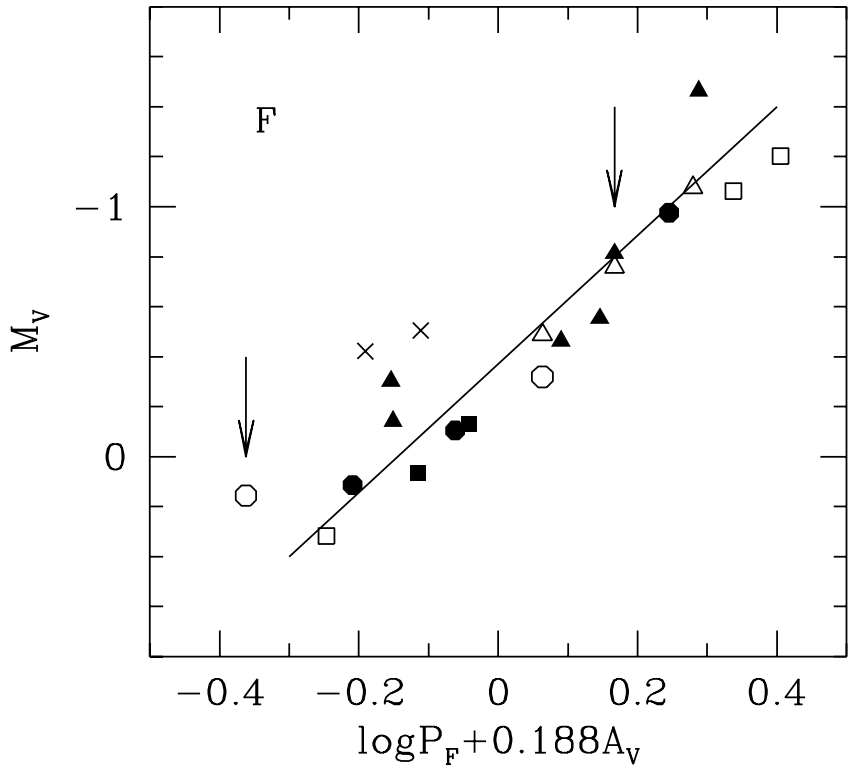
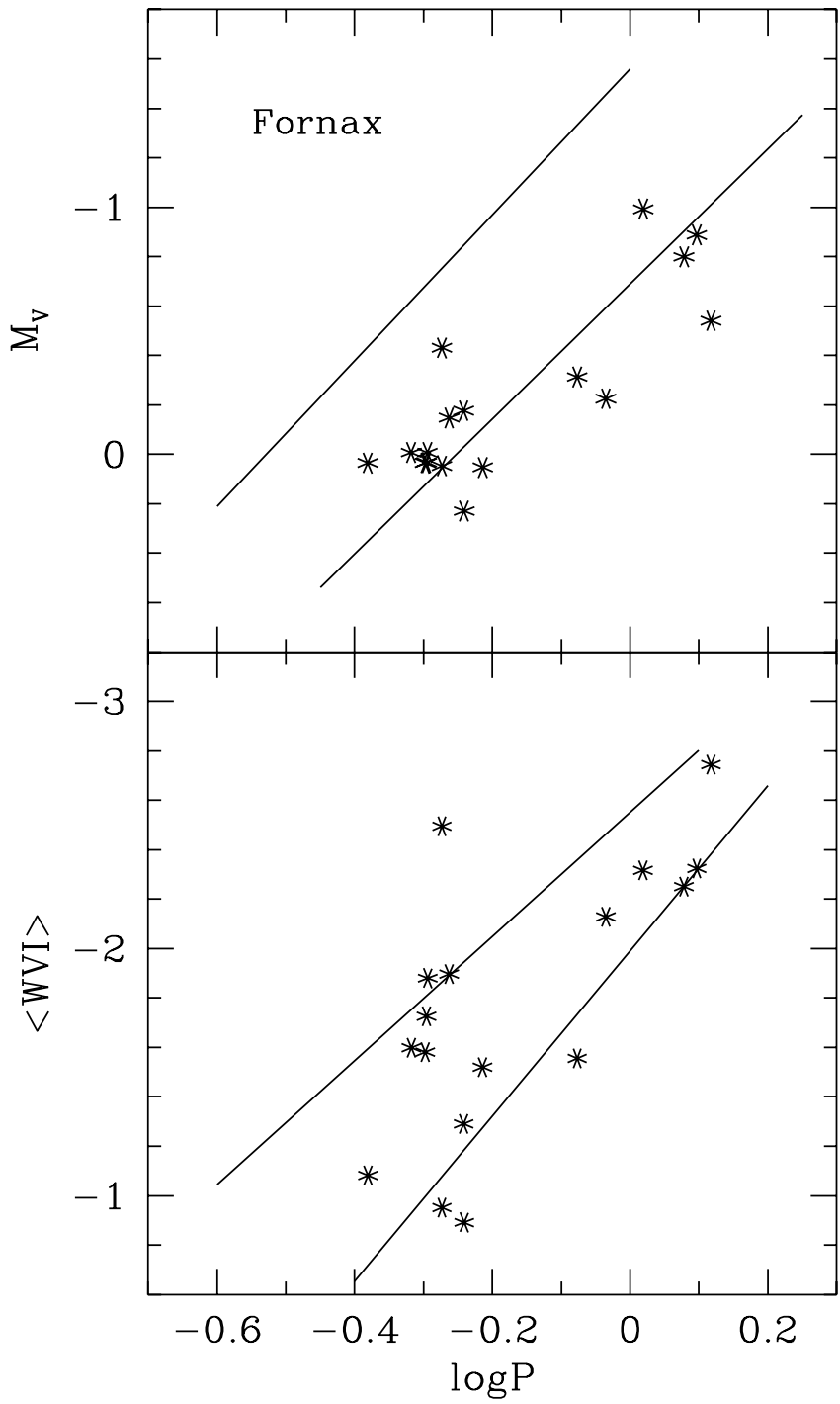
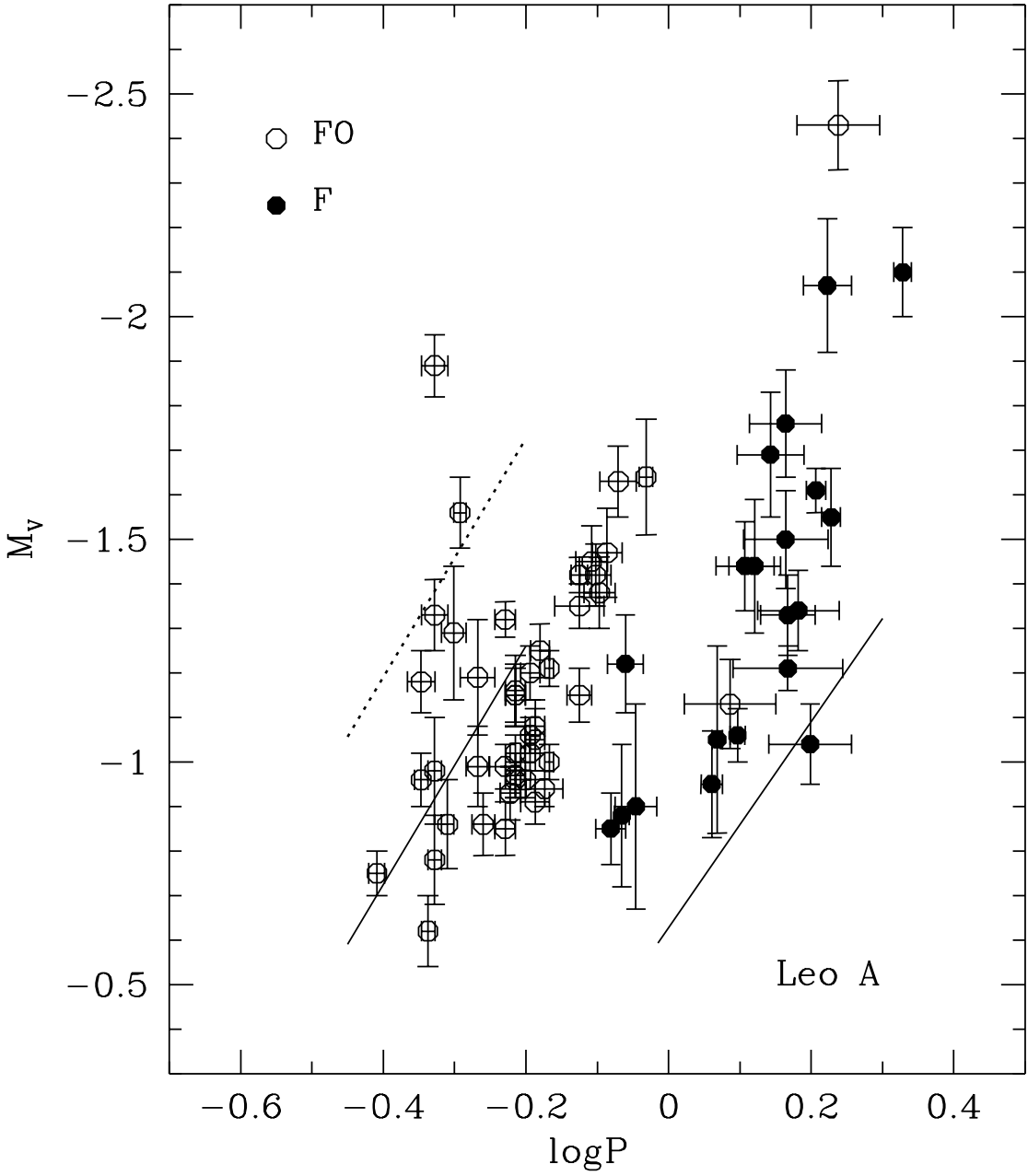
\end{document}